\documentclass[twocolumn]{aastex631}
\usepackage{mathtools} 
\usepackage{csvsimple}
\usepackage{longtable}
\usepackage{hyperref}

\usepackage{xcolor}

\usepackage{placeins}

\shorttitle{Minute-Scale Eclipse Regime in \textit{TESS}}
\shortauthors{Cavalier, de Wit, et al.}
\graphicspath{{./}{figures/}}
\usepackage{orcidlink}

\begin{document}


\title{A Previously Underexplored Regime in \textit{TESS}: \\
Minute-Scale Eclipses Reveal Ten White Dwarf--Cool M-Dwarf Binaries}

\correspondingauthor{T. Cavalier, J. de Wit}
\email{tcava@mit.edu, jdewit@mit.edu}

\author[0009-0001-4145-8929]{Tristan Cavalier}
\affiliation{Department of Earth, Atmospheric and Planetary Sciences, Massachusetts Institute of Technology, Cambridge, MA 02139, USA}
\affiliation{Institut Supérieur de l’Aéronautique et de l’Espace (ISAE-SUPAERO), Toulouse, 31055, France}

\author[0000-0003-2415-2191]{Julien de Wit}
\affiliation{Department of Earth, Atmospheric and Planetary Sciences, Massachusetts Institute of Technology, Cambridge, MA 02139, USA}

\author[0000-0001-9087-1245]{Mathilde Timmermans }
\affiliation{School of Physics \& Astronomy, University of Birmingham, Edgbaston, Birmingham B15 2TT, UK.}
\affiliation{Astrobiology Research Unit, Université de Liège, 19C Allée du 6 Août, 4000 Liège, Belgium}

\author[0000-0002-3627-1676]{Benjamin V.\ Rackham}
\affiliation{Department of Earth, Atmospheric and Planetary Sciences, Massachusetts Institute of Technology, Cambridge, MA 02139, USA}
\affiliation{Kavli Institute for Astrophysics and Space Research, Massachusetts Institute of Technology, Cambridge, MA 02139, USA}

\author[0009-0008-1188-2025]{Mariangel Albornoz}
\affiliation{Department of Earth, Atmospheric and Planetary Sciences, Massachusetts Institute of Technology, Cambridge, MA 02139, USA}
\affiliation{Department of Astronomy, University of Florida, Gainesville, FL 32611, USA}

\author[0000-0001-9892-2406]{Artem~Y.~Burdanov}
\affiliation{Department of Earth, Atmospheric and Planetary Sciences, Massachusetts Institute of Technology, Cambridge, MA 02139, USA}

\author[0000-0003-1464-9276]{Khalid Barkaoui }
\affiliation{Instituto de Astrofísica de Canarias (IAC), Calle Vía Láctea s/n, 38200, La Laguna, Tenerife, Spain.}
\affiliation{Department of Earth, Atmospheric and Planetary Sciences, Massachusetts Institute of Technology, Cambridge, MA 02139, USA}
\affiliation{Astrobiology Research Unit, Université de Liège, 19C Allée du 6 Août, 4000 Liège, Belgium}

\author[0000-0002-5510-8751]{Amaury H.M.J. Triaud}
\affiliation{School of Physics \& Astronomy, University of Birmingham, Edgbaston, Birmingham B15 2TT, United Kingdom}

\author[0000-0002-6523-9536]{Adam J.\ Burgasser}
\affiliation{Department of Astronomy \& Astrophysics, University of California San Diego, La Jolla, CA 92093, USA}

\author[0000-0002-9355-5165]{Brice-Olivier Demory}
\affiliation{Centerfor Space and Habitability, University of Bern, Gesellschaftsstrasse
6, 3012, Bern, Switzerland}

\author[0000-0002-7008-6888]{Elsa Ducrot}
\affiliation{AIM, CEA, CNRS, Université Paris-Saclay, Université de Paris,
F-91191 Gif-sur-Yvette, France}

\author[0000-0003-1462-7739]{Michael Gillon}
\affiliation{Astrobiology Research Unit, Université de Liège, 19C Allée du 6 Août, 4000 Liège, Belgium}

\author[0000-0002-7486-6726]{Yilen Gómez Maqueo Chew}
\affiliation{Universidad Nacional Autónoma de México, Instituto de As-
tronomía, AP 70-264, Ciudad de México, 04510, México}

\author[0000-0003-0030-332X]{Matthew J. Hooton}
\affiliation{Cavendish Laboratory, JJ Thomson Avenue, Cambridge CB3 0HE, UK}

\author[0000-0002-5220-609X]{Peter P. Pedersen}
\affiliation{Cavendish Laboratory, JJ Thomson Avenue, Cambridge CB3 0HE, UK}
\affiliation{Institute for Particle Physics and Astrophysics, ETH Zürich,
Wolfgang-Pauli-Strasse 2, 8093 Zürich, Switzerland}

\author[0000-0002-3012-0316]{Didier Queloz}
\affiliation{Cavendish Laboratory, JJ Thomson Avenue, Cambridge CB3 0HE, UK}
\affiliation{Institute for Particle Physics and Astrophysics, ETH Zürich,
Wolfgang-Pauli-Strasse 2, 8093 Zürich, Switzerland}

\author[0000-0002-9350-830X]{Sebastián Zúñiga-Fernández}
\affiliation{Astrobiology Research Unit, Université de Liège, 19C Allée du 6 Août, 4000 Liège, Belgium}

\begin{abstract}
Short-period white-dwarf (WD) binaries are post-common-envelope systems that constrain orbital shrinkage, envelope evolution, and the survival of low-mass companions. We report the discovery and confirmation of ten fully eclipsing short-period WD + cool M-dwarf binaries identified through a tailored search for minute-scale eclipses in \textit{Transiting Exoplanet Survey Satellite} (\textit{TESS}) high-cadence data and validated with \textit{SPECULOOS} multi-band photometry. The systems have orbital periods of a few hours and companions with effective temperatures of $\sim2700$--$3400$ K.
These discoveries demonstrate that \textit{TESS} contains a previously underexplored population of compact WD binaries whose short-duration, high-frequency, and often diluted eclipse signals are not efficiently recovered by standard transit-search pipelines. Whereas the literature contained only one eclipsing WD+M binary reported as a \textit{TESS}-based discovery, our pilot search of $\sim3.7\times10^4$ \textit{Gaia}-selected WDs yields ten new confirmed systems, increasing the \textit{TESS}-discovered sample by an order of magnitude. \textit{SPECULOOS} follow-up confirms the eclipses occur on target and uses chromatic eclipse dilution to distinguish stellar from substellar companions.
We combine multi-band eclipse photometry with Bayesian spectral energy distribution modeling to derive self-consistent WD and companion parameters. The resulting systems expand the known population of fully eclipsing WD+M binaries and notably double the number of systems in temperature regimes corresponding to M4 and M7 companions. This work establishes a scalable framework for identifying compact WD binaries in time-domain photometric surveys. Applied to \textit{TESS} archival data across the full \textit{Gaia} WD-candidate catalog ($\sim1.3\times10^6$ sources), and eventually to future surveys, this approach opens the prospect of assembling a population large enough to constrain post-common-envelope evolution and the stellar--substellar transition.

\end{abstract}


\section{Introduction}
\label{sec:intro}

Binary systems, in which two stars orbit a common center of mass, are ubiquitous across the Milky Way. Their occurrence rises steeply with primary mass, from roughly 40\% for solar-type stars to nearly 90\% for O-type main-sequence (MS) stars \citep{SatMoe2017}. 

A subset of these binaries evolves at sufficiently small separations to undergo mass transfer and/or a common-envelope (CE) phase \citep{Paczynski1976}, the physics of which remains only partially understood. Binary systems in which a white dwarf (WD) is orbited by a close companion are expected to form through a CE episode, during which the envelope of the WD progenitor is ejected \citep{Ivanova2013}. Post-common-envelope WD + main-sequence binaries (WDMS) therefore provide stringent empirical tests of binary evolution pathways and CE prescriptions \citep{Webbink2008,Ivanova2020}. Using model-independent measurements for 23 M dwarfs in WDMS, \citet{Parsons2018} found substantial object-to-object scatter, with 75\% of their sample showing radius inflation of up to 12\%. Consistently, \citet{Kesseli2018} suggested that fully convective M dwarfs may be systematically oversized relative to evolutionary models by approximately 13--18\% at the lowest masses, $0.08 < M < 0.18\,M_{\odot}$. Whether to identify population-level trends or to refine the initial and final conditions used in CE simulations, the discovery of new systems remains essential \citep{Moreno2022,Bronner2024,Green2025,Shariat2026}.

Over the past decade, large-scale sky surveys have identified a substantial population of WDMS, primarily through spectroscopy---see, e.g., SDSS \citep{York2000}, the Catalina Sky Survey \citep{Drake2009}, and ZTF \citep{Bellm2019}. \citet{Rebassa2021} estimated a WDMS space density of $3.7 \times 10^{-4}\,\mathrm{pc^{-3}}$, suggesting that roughly $\simeq 1500$ WDMS binaries should reside within 100\,pc. Within this broader population, eclipsing systems are particularly valuable, as their geometry enables precise measurements of orbital configurations and tight constraints on the present-day physical properties of both components, including uncontaminated spectra in fully eclipsing systems. The \textit{Transiting Exoplanet Survey Satellite} (\textit{TESS}; \citealt{Ricker2015}) provides an advantageous starting point for identifying eclipsing WD companions and triggering ground-based follow-up campaigns \citep[e.g.,][]{DeWit_WDBD_2025}. However, the short-duration, high-frequency, and often diluted eclipses expected for compact WD binaries occupy a regime that is not efficiently recovered by standard transit-search pipelines.

In this paper, we present the discovery and confirmation of ten short-period eclipsing post-common-envelope binaries (PCEBs) consisting of a WD primary and a low-mass M-dwarf companion. These systems represent the first stage of a broader survey effort aimed at identifying compact WD binaries across the full \textit{TESS} archive. Section~\ref{sec:target_selection_tess} describes the sample selection and search for eclipse signals. Section~\ref{sec:followup} details the ground-based multi-band photometric follow-up used to confirm that the detected transits occur on target and to initiate companion classification through chromatic transit dilution. Section~\ref{sec:sed_modeling} introduces the spectral energy distribution modeling used to further constrain the system parameters. We present the results in Section~\ref{sec:results}, discuss their implications in Section~\ref{sec:discussion}, and summarize our conclusions in Section~\ref{sec:conclusion}.


\begin{figure*}[!ht]
    \centering
    \includegraphics[trim = 0 20mm 0 10mm, width=0.9\textwidth]{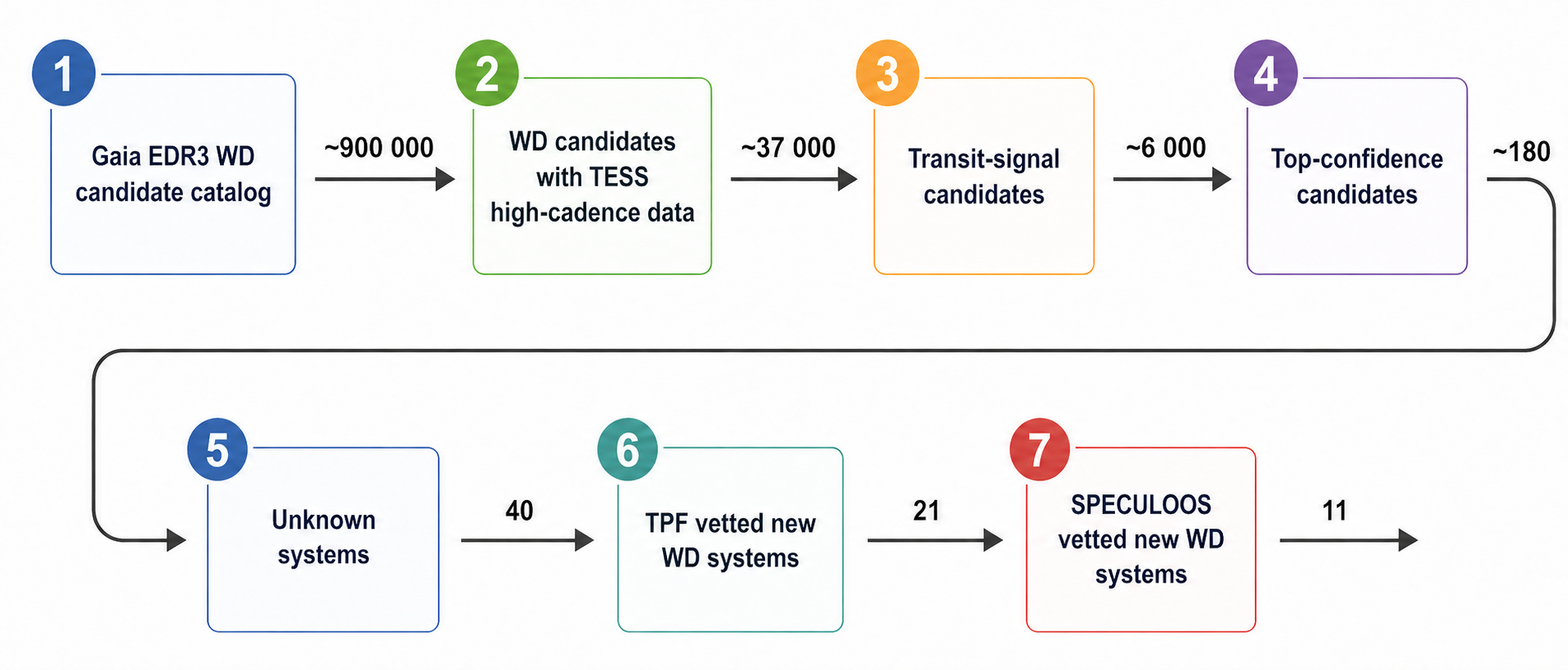}
    \caption{
    \textbf{Pipeline overview from the original sample of WD candidates to the newly-identified WD systems.} The pipeline takes in our inclusive \textit{Gaia} EDR3 WD-candidate catalog with $\sim$900,000 WD candidates, identifies $\sim$37,000 objects with \textit{TESS}' high-cadence data, $\sim$6,000 with signals flagging them as candidate eclipsing systems, $\sim$180 of them are high-SNR, of these a dozen are ultimately identified as new WD systems. TPF refers to \textit{TESS} Target Pixel Files, and \textit{SPECULOOS} denotes the ground-based photometric follow-up survey used for confirmation.
    }
    \label{fig:pipeline_overview}
\end{figure*}

\section{Target Selection and \textit{TESS} Detection}
\label{sec:target_selection_tess}

In this section, we describe the detection pipeline used to identify compact WD systems in \textit{TESS} high-cadence photometry. Starting from the \textit{Gaia} EDR3 WD-candidate catalog, we selected an intentionally inclusive sample of $\sim9\times10^5$ WD candidates, identified $\sim3.7\times10^4$ targets observed at 2-min cadence by \textit{TESS}, and searched their light curves for short-duration, high-frequency eclipses. This search yielded $\sim6000$ candidate signals, $\sim180$ high-priority candidates after initial vetting, and ultimately a dozen new compact WD systems following pixel-level validation and ground-based follow-up (Figure~\ref{fig:pipeline_overview}). As a validation check, the pipeline recovers known WD companions such as the WD planet candidate WD~1856+534~b \citep{Vanderburg2020} and the WD brown dwarf WD~0038+2030~B \citep{vanRoestel2021}.

\subsection{\textit{Gaia} EDR3 White Dwarf Sample}
\label{subsec:gaia_sample}

We constructed our parent sample from the \textit{Gaia} EDR3 white-dwarf candidate catalog \citep{Gentile2021}. We used the \textit{Gaia} EDR3 implementation of the white-dwarf probability, $P_{\mathrm{WD}}$, noting that the underlying probabilistic classifier was originally developed and validated on \textit{Gaia} DR2. We adopted an intentionally inclusive threshold of $P_{\mathrm{WD}} > 0.10$ to retain both high-confidence WDs and plausible candidates for subsequent vetting, yielding an initial sample of $\sim9\times10^5$ objects.

We then cross-matched this list with the \textit{TESS} short-cadence targets \citep{Stassun2019} and identified $\sim3.7\times10^4$ potential WDs monitored at 2-min cadence, defining the search sample analyzed here.

\subsection{Transit Detection Method}
\label{subsec:transit_detection}

We searched the 2-min \textit{TESS} light curves for short-duration flux decrements using a Box-Least-Squares (BLS) periodogram \citep{Kovacs2002}. We used the pre-search data conditioning simple aperture photometry (PDCSAP) light curves produced by the Science Processing Operations Center (SPOC) pipeline \citep{Jenkins2016}, which mitigate instrumental systematics and scattered-light contamination relative to the raw SAP fluxes.

The search was optimized for short and frequent eclipses, with periods spanning 0.1 hr to a few days. Although eclipsing WD+M binaries can produce deep events in high-quality photometry, \textit{TESS}'s large pixel scale ($\sim21^{\prime\prime}$) and the intrinsic faintness of WDs often dilute the signal and make detections challenging. These properties motivate a dedicated search and vetting strategy tailored to minute-scale eclipses.

Our pipeline identified $\sim6000$ candidate signals, which were reduced to $\sim180$ high-priority candidates after initial automated and visual vetting. Throughout this paper, we use TIC~450781262 as a representative example to illustrate the detection and validation procedure (Figure~\ref{fig:lc_tic450781262}). The full set of diagnostic plots for the confirmed systems is provided in Appendix~\ref{appendix:diagnostics}.

We then performed a literature search to identify systems already reported or characterized in previous work, reducing the sample to 40 new compact-WD candidates. To reject signals arising from nearby eclipsing binaries, planetary systems, or \textit{TESS} stray-light contamination, we performed pixel-level vetting using \textit{TESS} Target Pixel Files (TPFs). For each candidate and available sector with an exposure time shorter than 10 minutes, we downloaded the corresponding TPF, constructed a Gaussian PSF model to define a weighted pixel mask, and extracted a PSF-weighted light curve. We then computed a per-pixel SNR map from phase-folded light curves to test whether the signal was spatially consistent with the WD target (Figure~\ref{fig:tpf_tic450781262}).

This additional vetting rejected approximately half of the new candidates, including four newly identified background eclipsing binaries and 15 weak or spatially inconsistent signals, such as events associated with systematics shared across multiple pixels. The remaining candidates were prioritized for ground-based follow-up with \textit{SPECULOOS}.

\begin{figure*}[!ht]
    \centering
    \includegraphics[trim = 0 5mm 0 0, width=0.95\textwidth]{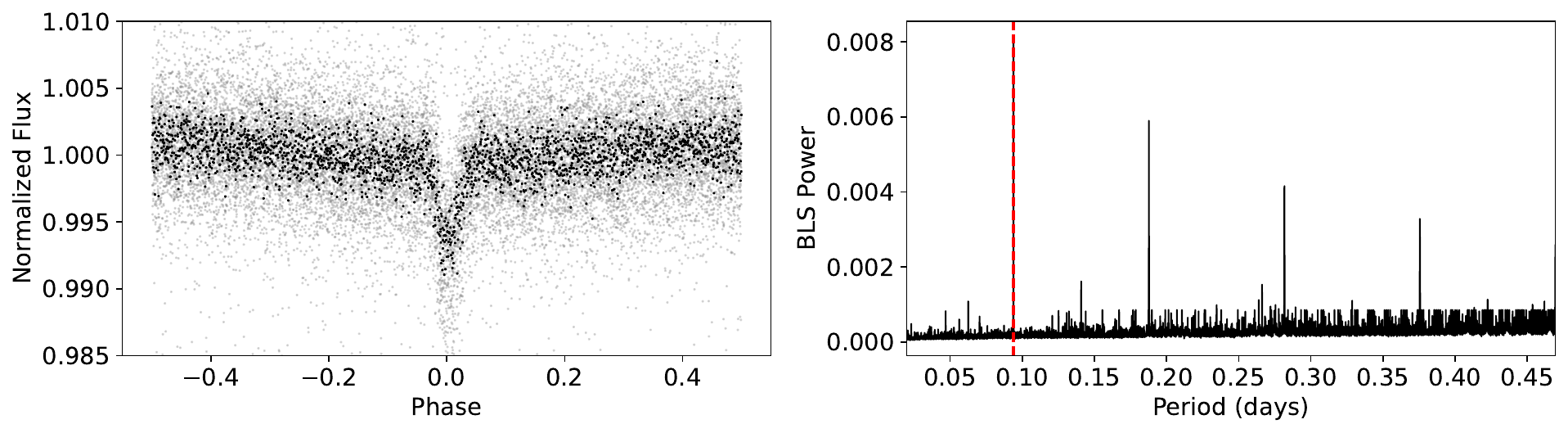}
    \caption{
    \textbf{Detection example for  narrow and frequent transit signals in \textit{TESS} high-cadence data.} \textit{Left:} Phase-folded light-curve of TIC450781262 for all available \textit{TESS} sectors showing a periodic $\sim$ 7-min transit event. The shallow depth (0.5\%) is due to the dilution of the full transit by bright nearby stars (see Fig. 3). \textit{Right:} Box-Least-Squares (BLS) periodogram of TIC450781262's \textit{TESS} light-curve revealing the orbital period of the system: P$\sim$2.25\,hrs.
    }

    \label{fig:lc_tic450781262}
\end{figure*}

\begin{figure*}[!htbp]
    \centering
    \includegraphics[width=\textwidth]{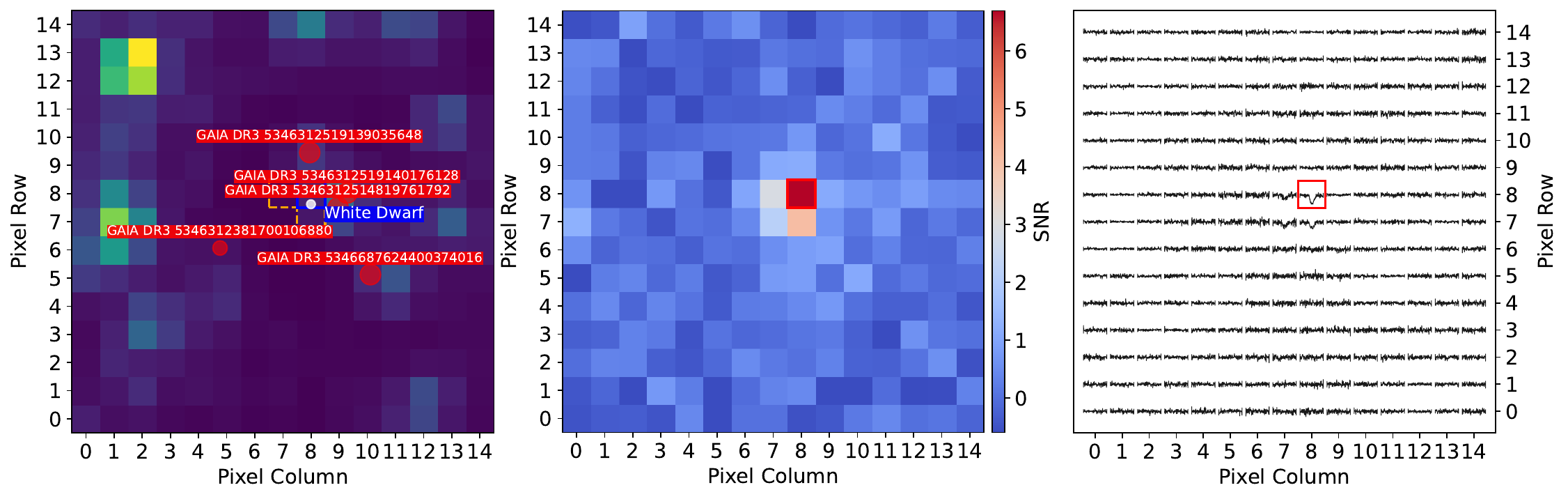}
    \caption{
    \textbf{Pixel-level vetting of TIC450781262 in \textit{TESS} Sector 99.}
    \emph{Left:} median 15$\times$15 Target Pixel File (TPF) cutout showing the WD target (white marker) and nearby \textit{Gaia} DR3 sources (red markers).
    \emph{Middle:} pixel-by-pixel signal-to-noise ratio (SNR) map of the transit-like signal, computed from phase-folded light curves, with the highest-SNR pixel highlighted.
    \emph{Right:} grid of phase-folded and binned light curves for all pixels in the cutout; the pixel showing the strongest transit-like signature is boxed in red.
    The spatial coincidence between the SNR peak and the target position supports an on-target origin of the signal rather than contamination from a nearby source.
    }
    \label{fig:tpf_tic450781262}
\end{figure*}

\pagebreak

\newpage

\section{Follow-up with \textit{SPECULOOS}}
\label{sec:followup}

\subsection{Confirming on-target eclipses}

We used the \textit{SPECULOOS} network of 1-m class telescopes to confirm that the candidate eclipse signals remaining after TPF-based vetting originate from the target WDs \citep{Delrez2018,Sebastian2021,Burdanov2022}. \textit{SPECULOOS} provides substantially higher angular resolution than \textit{TESS} ($0.35^{\prime\prime}$ pixel$^{-1}$ versus $\sim21^{\prime\prime}$ pixel$^{-1}$), reducing contamination from nearby sources. Of the 21 candidate systems for which the TPF data could not reliably identify the origin of the signal, \textit{SPECULOOS} confirmed 11 as on-target WD systems and identified 10 as background eclipsing binaries, including an extremely short-period binary composed of two late-M dwarfs (Cavalier et al., in prep.). The confirmed WD sample comprises the ten WDMS systems presented here and one WD+BD system, the low-mass, ultra-hot WD~J0404+1112 \citep{DeWit_WDBD_2025}.

\subsection{Dilution-based characterization}

We then leveraged the multi-band capability of \textit{SPECULOOS} to search for chromatic variations in eclipse depth, providing first constraints on the temperature and nature of the companions. Because WDs are intrinsically faint in the near-infrared (NIR), while late-type stellar companions emit more strongly at these wavelengths, eclipses produced by stellar companions appear shallower in redder filters owing to flux dilution. In contrast, cooler brown-dwarf companions contribute negligibly at optical/NIR wavelengths and therefore produce nearly achromatic, near-total eclipses. Using broad optical-to-NIR filters (Exo\footnote{A blue-blocking filter with transmittance $>90\%$ from $\sim$500 nm to $>1000$ nm.}, Sloan $i'$, and $I+z$) and achieving 0.5--2\% photometric precision, we detect significant eclipse-depth variations across bandpasses. These chromatic signals allow us to distinguish stellar from substellar companions and, in many cases, rule out cooler brown dwarfs in favor of cool-M-dwarf companions.

We modeled the measured eclipse depths using synthetic band-integrated flux ratios computed from atmosphere models. For the WD flux, we used Koester DA atmosphere models \citep{Bedard2020}, with the WD effective temperature \(T_1\), radius \(R_1\), and surface gravity \(\log g_1\) inferred from the SED and cooling-track analysis described in Section~\ref{sec:sed_modeling}. The companion flux was modeled using the PHOENIX atmosphere grid \citep{Husser2013} over a range of effective temperatures \(T_2\), implemented through the \texttt{speclib} Python package \citep{speclib,Rackham2024}. For each model, we integrated the synthetic spectra through the corresponding \textit{SPECULOOS} transmission curve to obtain the band-averaged surface fluxes \(F_{1,b}\) and \(F_{2,b}\). The companion-to-WD flux ratio in band \(b\) is then

\[
q_b(T_2)=\left(\frac{R_2(T_2)}{R_1}\right)^2
\frac{F_{2,b}(T_2)}{F_{1,b}(T_1)}.
\]

The companion temperature \(T_2\) is sampled at each Monte Carlo iteration according to the dilution likelihood, while the corresponding radius \(R_2\) is assigned self-consistently as a function of \(T_2\) using the M-dwarf evolutionary models of \citet{Baraffe2015}.

Because the WD radius is much smaller than that of a M companion (\(R_1 \ll R_2\)), the eclipse of the WD is expected to be total for the systems considered here. The observed fractional eclipse depth in band \(b\) is therefore set by the WD contribution to the total system flux and is given by

\[
\Delta_b = \frac{1}{1+q_b}.
\]

\begin{figure}[ht]
    \centering
    \includegraphics[width=\columnwidth]{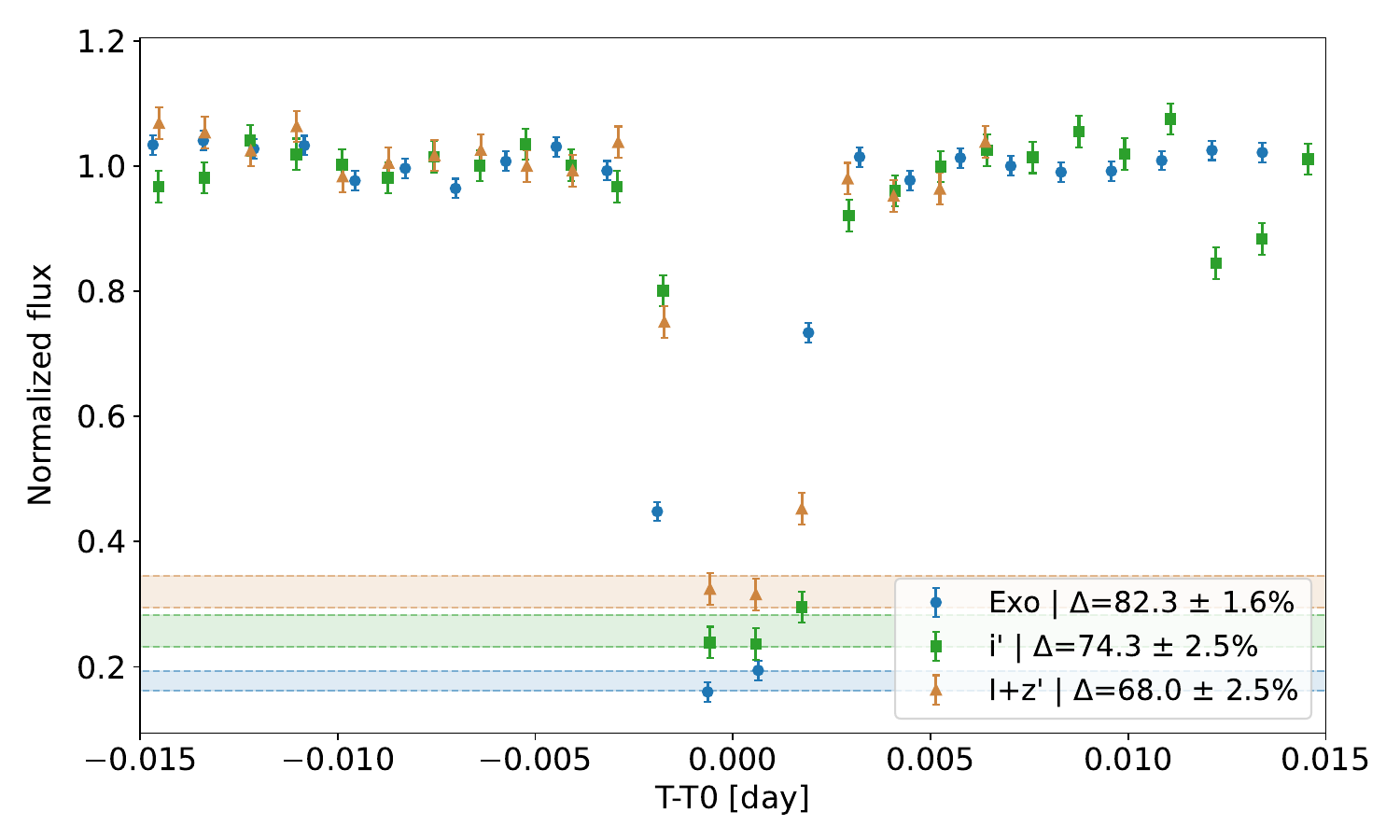}
    \caption{
    \textbf{\textit{SPECULOOS}' follow-up observations of TIC~450781262.}
    Follow-up observations of TIC~450781262 obtained with the \textit{SPECULOOS} Northern Observatory (SNO) with the telescope Artemis in the I+z (orange), i’ (green), and Exo (blue) filters confirms that the transit signals detected in \textit{TESS} are on-target, while the transit duration (corresponding to a Jupiter size object) and the transit-depth chromaticity (corresponding to a $\sim$3,000\,K object) reveal the nature of the companion: a cool-M dwarf.}

    \label{fig:lc_spec_tic450781262}
\end{figure}


\section{Spectral Energy Distribution Modeling}
\label{sec:sed_modeling}

The spectral energy distribution (SED) provides a complementary constraint on the presence and nature of a cool companion by exploiting the strong temperature contrast between a hot white dwarf and a late-type object. While the WD dominates the emission at blue-optical wavelengths, a cool companion contributes increasingly in the red and near-infrared, such that a systematic infrared excess signals an additional emitter.

The SED photometry was assembled using the CDS VizieR SED tool \citep{Ochsenbein2000}, combining archival measurements from major optical and infrared surveys, including \textit{Gaia} DR3 \citep{GaiaDR3}, the \textit{Sloan Digital Sky Survey} (SDSS; \citealt{SDSS}), the \textit{Two Micron All Sky Survey} (2MASS; \citealt{2MASS}), the \textit{Pan-STARRS1} surveys \citep{Pan-STARRS1}, and the \textit{AllWISE} data release \citep{AllWISE}.

We analyzed the SEDs using a Bayesian nested-sampling framework implemented with \texttt{UltraNest} \citep{Buchner2021}, enabling exploration of parameter degeneracies and a statistical comparison between single- and two-component models. The assembled multi-survey photometry was first homogenized by removing duplicate measurements and excluding data points inconsistent with the local continuum. The SED normalization was parameterized through the scaling factor $(R/d)^2$, with a log-uniform prior on the WD radius, \(R_1\), and a Gaussian prior on the distance derived from the \textit{Gaia} DR3 parallax. This approach reduces the intrinsic radius--distance degeneracy and anchors the inferred radii to the astrometric distance.

\begin{figure}[!htbp]
    \centering
    \includegraphics[width=\columnwidth]{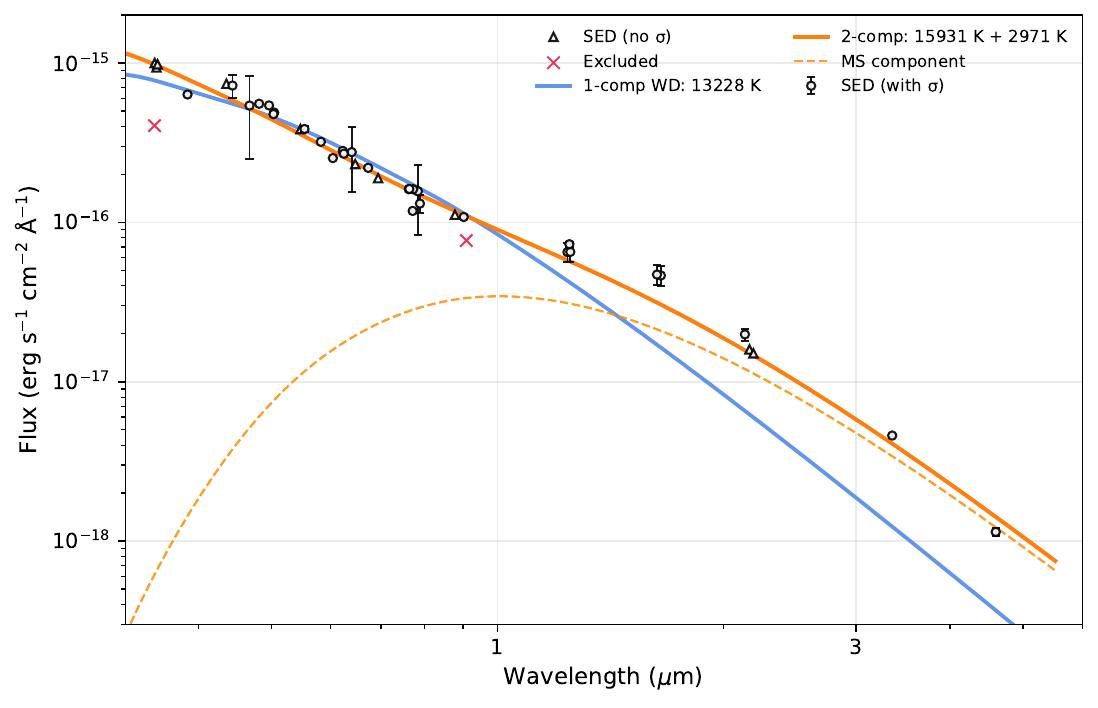}
    \caption{
    \textbf{Spectral Energy Distribution (SED) of TIC~450781262.} An analysis of the TIC~450781262's SED best explains the flux excess beyond 1 microns as originating from a $\sim$ 3,000\,K cool-M companion. 
    }
    \label{fig:sed_tic450781262}
\end{figure}

The analysis was formulated as a model-comparison problem between a single hot component and a composite hot+cold solution, testing whether the observed flux distribution requires an additional cool emitter to reproduce the red or near-infrared excess (Figure~\ref{fig:sed_tic450781262}). For the confirmed systems, the two-component model is strongly favored, providing clear evidence for cool companions consistent with WD+cool-M binary configurations.

Broadband SED fitting primarily constrains the global continuum shape and flux normalization, and is therefore most sensitive to temperature contrast and infrared excess rather than detailed atmospheric properties. The inferred parameters are therefore interpreted jointly with the independent constraints from eclipse photometry and spectroscopy. The corresponding SED fits and posterior distributions for each system are shown in Appendix~\ref{appendix:diagnostics}.

\begin{table}[ht]
\centering
\caption{Best-fit SED parameters for TIC450781262.}
\label{tab:sed_fit_results}
\begin{tabular}{lcc}
\hline
Parameter & 1-component & 2-component \\
\hline
$T_1$ (K) & $13228^{+349}_{-366}$ & $15931^{+3226}_{-1756}$ \\
$R_1$ ($R_\odot$) & $0.0309^{+0.0012}_{-0.0011}$ & $0.0203^{+0.0014}_{-0.0017}$ \\
$T_2$ (K) & -- & $2971^{+97}_{-95}$ \\
$R_2$ ($R_\odot$) & -- & $0.1479^{+0.0063}_{-0.0060}$ \\
$d$ (pc) & $296.7^{+6.4}_{-6.9}$ & $296.6^{+7.1}_{-6.5}$ \\
$A_V$ (mag) & $0.587^{+0.010}_{-0.020}$ & $0.242^{+0.216}_{-0.168}$ \\
\hline
$\log Z$ & $922.29 \pm 0.26$ & $1370.64 \pm 0.23$ \\
\hline
\end{tabular}
\end{table}

\section{Results}
\label{sec:results}

\begin{table*}[ht]
\centering
\caption{Orbital period, white dwarf effective temperature and radius, and the companion effective temperature and radius for each target derived from the joint fit of the SED and the transit dilution. }
\label{tab:joint_fit_parameters}

\small
\setlength{\tabcolsep}{6pt}
\renewcommand{\arraystretch}{1.2}

\begin{tabular}{lcccccc}
\hline
\hline

TIC ID &
$P$ (d) &
$T_0$ (BTJD) &
$T_{\rm WD}$ (K) &
$R_{\rm WD}$ ($R_{\odot}$) &
$T_{\rm comp}$ (K) &
$R_{\rm comp}$ ($R_{\odot}$) \\

\hline
450781262 &
$0.09388575 \pm 6.2\times10^{-7}$ &
$4005.84629 \pm 0.00064$ &
$16909_{-2246}^{+2614}$ &
$0.0200_{-0.0014}^{+0.0013}$ &
$2994_{-29}^{+14}$ &
$0.1475_{-0.0041}^{+0.0043}$ \\

53206761 &
$0.13280937 \pm 4.5\times10^{-8}$ &
$4067.61391 \pm 0.00082$ &
$26524_{-5688}^{+4598}$ &
$0.0157_{-0.0016}^{+0.0018}$ &
$3100_{-15}^{+5}$ &
$0.1905_{-0.0099}^{+0.0093}$ \\

2041210548 &
$0.29092036 \pm 2.6\times10^{-7}$ &
$3969.45289 \pm 0.00092$ &
$17258_{-1897}^{+3549}$ &
$0.0200_{-0.0018}^{+0.0016}$ &
$2764_{-58}^{+36}$ &
$0.1412_{-0.0073}^{+0.0076}$ \\

1000635787 &
$0.19178974 \pm 6.5\times10^{-7}$ &
$4154.43077 \pm 0.00090$ &
$36845_{-10682}^{+13841}$ &
$0.0109_{-0.0016}^{+0.0021}$ &
$3217_{-66}^{+63}$ &
$0.1656_{-0.0141}^{+0.0152}$ \\

1883504789 &
$0.15744210 \pm 9.9\times10^{-8}$ &
$3970.38020 \pm 0.00072$ &
$9978_{-180}^{+174}$ &
$0.0192_{-0.0007}^{+0.0007}$ &
$2699_{-25}^{+27}$ &
$0.0969_{-0.0038}^{+0.0039}$ \\

610682128 &
$0.09395693 \pm 1.0\times10^{-7}$ &
$3980.50103 \pm 0.00075$ &
$11459_{-309}^{+316}$ &
$0.0160_{-0.0012}^{+0.0012}$ &
$3187_{-43}^{+35}$ &
$0.1342_{-0.0104}^{+0.0096}$ \\

600854869 &
$0.20269374 \pm 2.5\times10^{-7}$ &
$4067.36917 \pm 0.00041$ &
$23097_{-933}^{+1101}$ &
$0.0258_{-0.0037}^{+0.0038}$ &
$2906_{-126}^{+99}$ &
$0.2079_{-0.0299}^{+0.0278}$ \\

1201294971 &
$0.28602015 \pm 3.0\times10^{-7}$ &
$4089.77399 \pm 0.00089$ &
$40525_{-12936}^{+10936}$ &
$0.0191_{-0.0024}^{+0.0043}$ &
$3004_{-47}^{+52}$ &
$0.1831_{-0.0114}^{+0.0133}$ \\

1936725512 &
$0.11596915 \pm 2.5\times10^{-7}$ &
$3979.44145 \pm 0.00095$ &
$35506_{-10558}^{+16699}$ &
$0.0207_{-0.0044}^{+0.0047}$ &
$3301_{-79}^{+97}$ &
$0.2511_{-0.0439}^{+0.0459}$ \\

1506668467 &
$0.20333078 \pm 4.6\times10^{-7}$ &
$4159.61721 \pm 0.00115$ &
$49542_{-4360}^{+7036}$ &
$0.0132_{-0.0004}^{+0.0003}$ &
$3368_{-4}^{+9}$ &
$0.1849_{-0.0081}^{+0.0051}$ \\

\hline
\end{tabular}
\end{table*}


The SED and transit-dilution analyses independently provide first-order constraints on the system components, but they probe complementary observables that depend on the same underlying physical parameters. To exploit this complementarity and avoid combining independent marginal estimates a posteriori, we performed a joint Bayesian fit of the SED and multi-band eclipse-depth measurements within a single posterior framework.

\begin{figure}[!ht]
    \centering
    \vspace{-0mm}\includegraphics[width=\columnwidth]{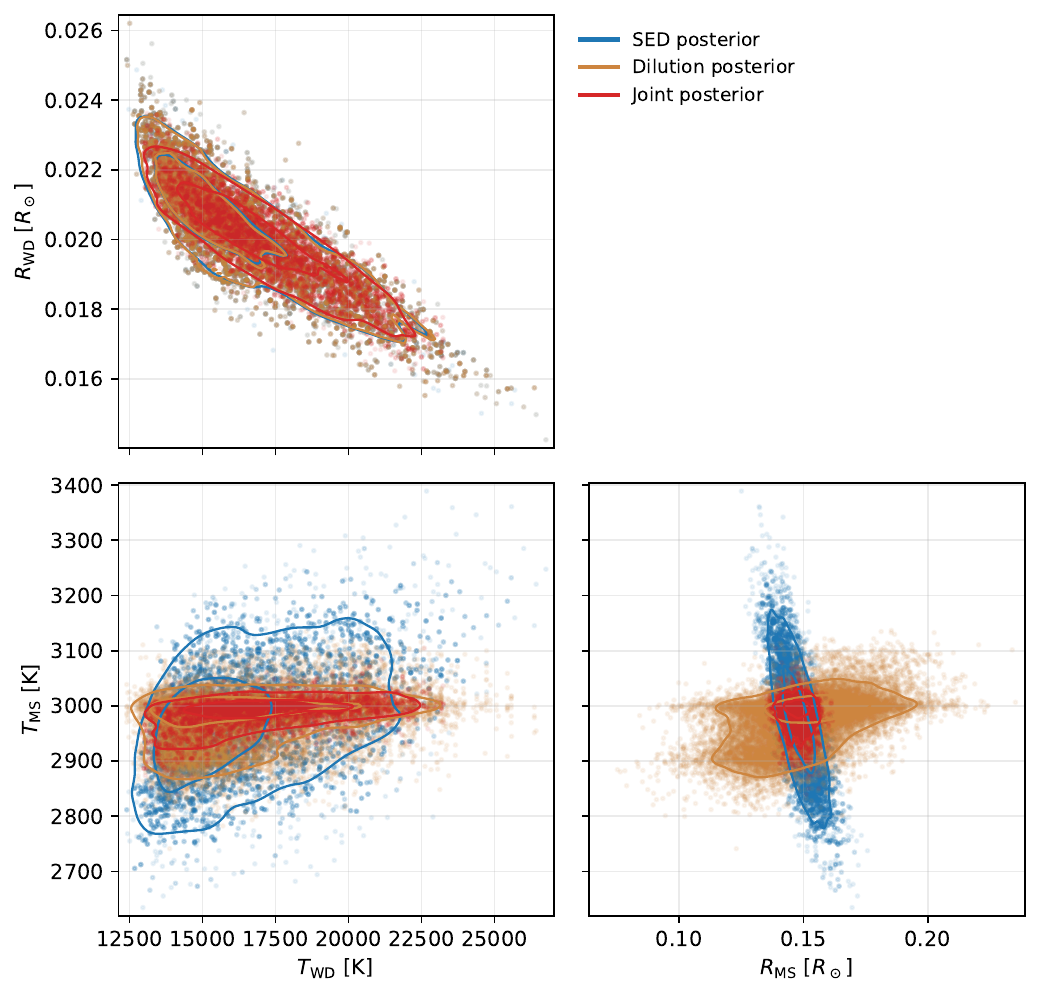}
    \caption{
    \textbf{Inferences on TIC~450781262's system.} 
    Posterior probability of TIC~450781262's key system parameters from transit-only (orange), SED-only (blue), and joint (red) fits showing good agreement between the complementary datasets.
    }
    \label{fig:joint_fit_corner}
\end{figure}

The resulting posterior distributions provide self-consistent estimates of $T_{\rm WD}$, $R_{\rm WD}$, $T_{\rm comp}$, and $R_{\rm comp}$, while naturally retaining the correlations between parameters. Figure~\ref{fig:joint_fit_corner} illustrates this improvement for TIC~450781262 by comparing the independent SED-only and dilution-only constraints with the joint posterior. The agreement between the independent constraints provides an important consistency check on the inferred binary configuration.

A summary of the derived physical parameters for all confirmed systems is presented in Table~\ref{tab:joint_fit_parameters}. The corresponding SED fits, multi-band eclipse measurements, and posterior distributions for each target are provided in Appendix~\ref{appendix:diagnostics}, following the same methodology illustrated for TIC~450781262.

For TIC~450781262, the one-component SED model requires a substantially larger extinction than the two-component model. This likely reflects the inability of a single hot component to reproduce the observed infrared excess: in the absence of a cool companion, the fit can partially compensate by adjusting both the WD temperature and extinction to reproduce the overall continuum shape. This behavior further supports the need for a two-component WD+cool-M interpretation.

\newpage
\subsection{Individual System Highlights}
\label{subsec:individual_systems}

The confirmed systems share the same broad phenomenology: a short-period eclipse signal in \textit{TESS}, confirmation of an on-target eclipse with \textit{SPECULOOS}, chromatic eclipse-depth variations consistent with dilution by a cool stellar companion, and an infrared excess in the broadband SED. The joint fits yield companion temperatures of $\sim2700$--$3400$~K and radii consistent with cool-M dwarfs, placing the sample near the hydrogen-burning limit.

TIC~450781262 (Figure~\ref{fig:Fig4507}) provides a representative benchmark case. Its SED shows a clear infrared excess beyond $1\,\mu$m, and the \textit{SPECULOOS} eclipse-depth chromaticity independently supports the presence of a M dwarf companion. TIC~53206761 (Figure~\ref{fig:Fig5320}), TIC~2041210548 (Figure~\ref{fig:Fig2041}), and TIC~1000635787 (Figure~\ref{fig:Fig1000}) show similarly consistent constraints from the SED and dilution analyses, leading to well-defined joint posteriors.

TIC~1883504789 (Figure~\ref{fig:Fig1883}), TIC~610682128 (Figure~\ref{fig:Fig6106}), and TIC~600854869 (Figure~\ref{fig:Fig6008}) also show coherent agreement between the SED and eclipse-depth constraints, although the inferred companion radii differ modestly between the two independent methods. The joint framework accounts for these differences by propagating the shared parameter correlations into the final posterior estimates.

TIC~1201294971 (Figure~\ref{fig:Fig1201}), TIC~1936725512 (Figure~\ref{fig:Fig1936}), and TIC~1506668467 (Figure~\ref{fig:Fig1506}) are systems with hotter WD primaries and cool late-type companions, but their SED constraints are weaker because of limited photometric coverage below $0.4\,\mu$m and between $1$ and $3\,\mu$m. In these cases, the infrared excess beyond $3\,\mu$m provides the main SED evidence for the companion, and improved UV/optical photometry would better constrain the WD parameters.

For the systems with weaker SED constraints, the consistency between the eclipse-depth and SED-based companion estimates, together with the \textit{SPECULOOS} on-target confirmations, argues against contamination by unresolved background eclipsing binaries. Archival imaging and \textit{Gaia} DR3 proper motions further show no detached M-dwarf source at the present-day WD positions. TIC~600854869, TIC~2041210548, and TIC~1883504789 were previously reported as WDMS candidates by \citet{Kosakowski2022} based on ZTF data.
\


\section{Discussion and Future Prospects}
\label{sec:discussion}

\subsection{Implications for Compact Binary Evolution}
\label{subsec:science_implications}

Short-period white-dwarf binaries are the outcome of a common-envelope (CE) phase, during which the orbit shrinks through the transfer of orbital energy and angular momentum to the envelope. Post-common-envelope systems therefore provide direct empirical constraints on the poorly understood physics of CE evolution. The ten fully eclipsing WD+cool-M systems presented here, with orbital periods of a few hours, probe this regime of strong orbital shrinkage and offer new benchmarks for testing CE prescriptions.

Their companions lie near the hydrogen-burning limit, bridging the populations of WD+stellar and WD+substellar systems. This regime provides important context for the apparent scarcity of close-in brown-dwarf companions, the ``brown dwarf desert'', by linking the initial companion-mass distribution to the population that survives the CE phase \citep{Grether2006,Chen2024}.

These systems are expected to evolve further through angular-momentum loss, including gravitational radiation and magnetic braking, eventually leading to Roche-lobe overflow and the formation of interacting binaries such as cataclysmic variables. They thus represent immediate progenitors of such systems, and long-term monitoring may reveal signatures of orbital decay.

\subsection{Context within the WDMS Population}
\label{subsec:population_context}

\begin{figure}[!ht]
    \centering
    \includegraphics[width=\columnwidth]{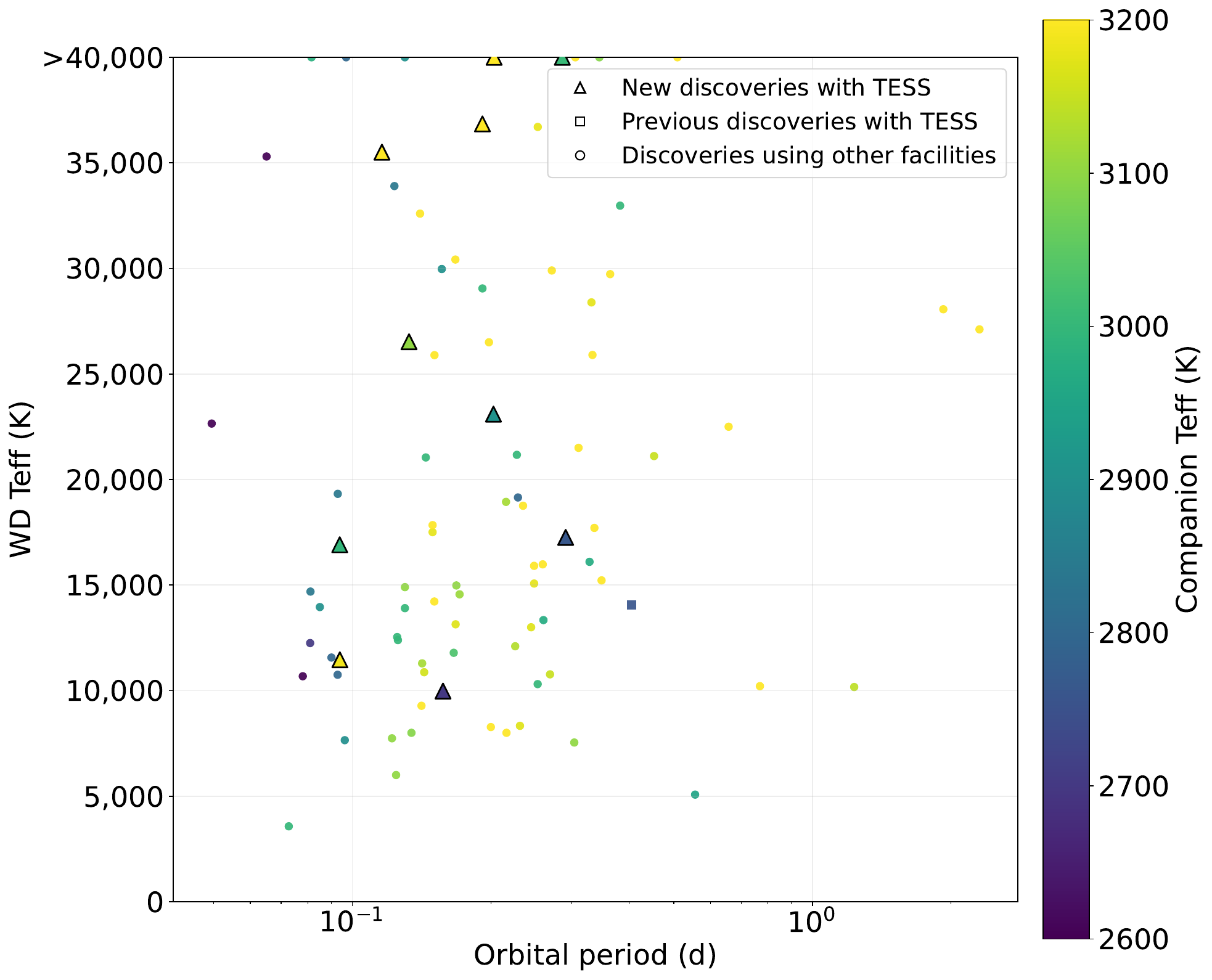}
    \caption{\textbf{New \textit{TESS} discoveries in the near-hydrogen-burning-limit WD+M population.} 
Known eclipsing WD+M systems are shown as a function of orbital period and white dwarf temperature, with companion temperature encoded by color. The single prior \textit{TESS}-based discovery \citep[TIC~60040774,][]{Priyatikanto2022} is shown as a square, while systems from other facilities are shown as circles. The ten new \textit{TESS} discoveries presented here (triangles) populate sparsely sampled companion-temperature regimes near the low-mass end of the M-dwarf sequence, notably doubling the number of known eclipsing WD+M systems in the temperature bins corresponding to M4 and M7 companions, demonstrating the added value of tailored searches for minute-scale eclipses in \textit{TESS} data.}
\label{fig:population}
\end{figure}

Fully eclipsing WD+M binaries are rare but uniquely informative, enabling model-independent constraints on both components. Figure~\ref{fig:population} places our systems within the known population in orbital period--white dwarf temperature space, with companion temperature encoded as a color scale.

Only $\sim70$--$80$ well-characterized eclipsing systems are currently known \citep[e.g.,][]{Parsons2015,Parsons2018,Rebassa2025}, primarily identified through spectroscopic and variability surveys and biased toward hotter white dwarfs and shorter periods. The ten systems presented here therefore represent a significant increase, particularly at the lowest companion masses.

Our sample extends the population toward cooler companions and into a sparsely explored regime near the hydrogen-burning limit. This region has remained observationally underconstrained despite its importance for binary evolution and the stellar--substellar transition. Because existing samples are shaped by heterogeneous selection functions, our approach---targeting minute-scale eclipses in high-cadence photometry---provides a complementary and more systematic view of this parameter space. These systems therefore help bridge the gap between WD+early-M binaries and WD+substellar companions.

\subsection{Prospects for Large-Scale Discovery and Follow-up}
\label{subsec:future_prospects}

Our results demonstrate that the \textit{TESS} archive contains a largely unexplored population of short-period white dwarf binaries accessible through tailored searches for minute-scale eclipses. Our pilot analysis of $\sim3.7\times10^{4}$ \textit{Gaia}-selected white dwarfs has already yielded ten fully eclipsing systems, highlighting the efficiency of this approach. Extending this methodology to the full \textit{Gaia} white dwarf catalog ($\sim1.3\times10^{6}$ sources) opens the prospect of identifying hundreds of compact binaries and placing population-level constraints on their occurrence rates and properties.

The \textit{TESS} extended mission, including 20-second cadence observations, will further improve sensitivity to ultra-short-duration eclipses. Future facilities such as \textit{PLATO} \citep{Rauer2014} and the \textit{Roman Space Telescope} \citep{Akeson2019} will provide complementary capabilities, extending this search to longer baselines, higher precision, and redder wavelengths.

\begin{figure}[!htbp]
    \centering
    \includegraphics[width=\columnwidth]{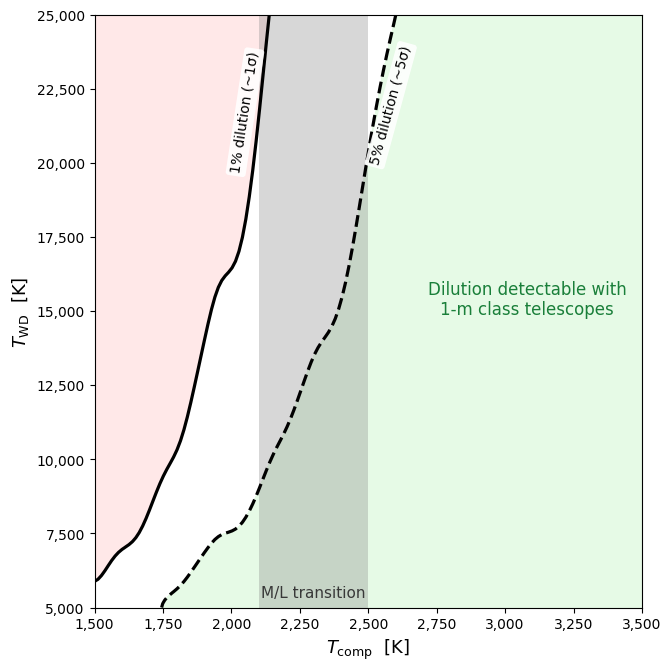}
    \caption{\textbf{Sensitivity of 1-m class telescopes to transit dilution in WD binaries.} 
Solid and dashed curves show the 1$\sigma$ and 5$\sigma$ detection thresholds for a single transit observation with a 1-m class telescope, corresponding to 1\% and 5\% transit dilution in the visible, respectively. The curves are shown as a function of white dwarf temperature ($T_{\rm WD}$) and companion temperature ($T_{\rm comp}$), with the shaded region indicating the transition between stellar and substellar companions. This illustrates that 1-m class facilities can provide a first-order discrimination between stellar and substellar companions through chromatic dilution, in addition to confirming that transit signals detected by surveys such as \textit{TESS} occur on target.}
\label{fig8}
\end{figure}

Ground-based follow-up is particularly well suited to short-period PCEBs, as multiple eclipse events can often be observed during a single night. As demonstrated here, 1-m class telescopes such as \textit{SPECULOOS} \citep{Delrez2018,Sebastian2021,Burdanov2022} are capable of confirming on-target eclipses and performing preliminary classification of compact WD systems. Their higher spatial resolution mitigates contamination, while multi-band photometry enables the identification of chromatic eclipse dilution, constraining companion temperatures and distinguishing stellar from substellar objects (Figure~\ref{fig8}). Additional narrow-band eclipse observations can further identify candidate systems hosting cyclotron-emitting magnetic WDs, as demonstrated for 2MASSJ0129+6715 by \citet{2020MNRAS.493.5208K}, enabling studies of magnetic-field strengths and their role in binary evolution \citep{2025A&A...696A.242V}.

The synergy between space-based surveys and modest-aperture ground-based networks thus provides a powerful framework for systematically exploring compact binary populations. The methodology validated here represents a first step, with ongoing analyses of the full \textit{TESS} dataset expected to substantially expand the known population in the near future.

\section{Conclusion}
\label{sec:conclusion}

We report the discovery and confirmation of ten fully eclipsing short-period white dwarf + cool-M dwarf binaries identified in \textit{TESS} high-cadence data and validated with \textit{SPECULOOS} multi-band photometry. With orbital periods of a few hours and companions near the hydrogen-burning limit, these systems probe a critical regime of post-common-envelope evolution and provide new benchmarks for constraining orbital shrinkage and the fate of low-mass companions.

These discoveries demonstrate that minute-scale eclipses in compact WD binaries define a previously underexplored discovery regime in \textit{TESS}. Whereas only one fully eclipsing WD+M binary had previously been reported as a \textit{TESS}-based discovery, the tailored search presented here yields ten new confirmed systems from a pilot sample of $\sim3.7\times10^4$ \textit{Gaia}-selected WDs. This order-of-magnitude increase highlights the value of searches optimized for short-duration, high-frequency, and often diluted eclipse signals.

By combining high-cadence space-based photometry with \textit{SPECULOOS} chromatic eclipse measurements and SED modeling, we derive self-consistent system parameters and demonstrate an efficient framework for identifying and characterizing compact binaries in large photometric surveys. The resulting sample expands the population of known fully eclipsing WD+cool-M systems, approximately doubles the number of known eclipsing WD+M systems in the companion-temperature bins corresponding to roughly M4 and M7 dwarfs, and occupies a previously underexplored region of parameter space linking stellar and substellar companions.

More broadly, this work establishes a scalable pathway toward population-level studies of compact WD binaries. Applied to archival \textit{TESS} data across the full \textit{Gaia} WD catalog, and eventually to future surveys, this approach opens the prospect of assembling a statistically significant sample of compact WD binaries with which to constrain post-common-envelope evolution, the stellar--substellar transition, and the long-term evolution toward interacting systems.

\section*{Acknowledgements}
J.d.W.\ and MIT gratefully acknowledge financial support from the Heising-Simons Foundation, Dr.\ and Mrs.\ Colin Masson and Dr.\ Peter A.\ Gilman for Artemis, the first telescope of the \textit{SPECULOOS} network situated in Tenerife, Spain. M.\ Gillon is FNRS Research Director. He and
ULiege thank the Wallonian Region for its contribution to the funding of the SPECULOOSNorth/Artemis telescope.
This material is based upon work supported by the National Aeronautics and Space Administration under Agreement No.\ 80NSSC21K0593 for the program ``Alien Earths.''
The results reported herein benefited from collaborations and/or information exchange within NASA's Nexus for Exoplanet System Science (NExSS) research coordination network sponsored by NASA's Science Mission Directorate.

The \textit{TESS} short-cadence light curves of the 10 new WDMS eclipsing systems used in this paper can be found in MAST : \dataset[10.17909/tm34-r693]{http://dx.doi.org/10.17909/tm34-r693}.

\bibliography{bibliography}{}
\bibliographystyle{aasjournal}

\clearpage

\appendix
\restartappendixnumbering

\section{Supplementary diagnostic plots}
\label{appendix:diagnostics}

This appendix presents the diagnostic plots (i.e., data and posterior probability distributions) used to derive the parameter values reported in Table\,\ref{tab:joint_fit_parameters}. Following the same methodology described throughout this work and illustrated using TIC~450781262 as a reference example, these figures provide additional support for the adopted physical parameters of the white dwarf and its companion, including the SED fitting results, posterior distributions, and the consistency between the inferred parameters and the observed multi-band eclipse depths. 

\begin{figure}[!htbp]
    \centering
    \includegraphics[width=1\textwidth]{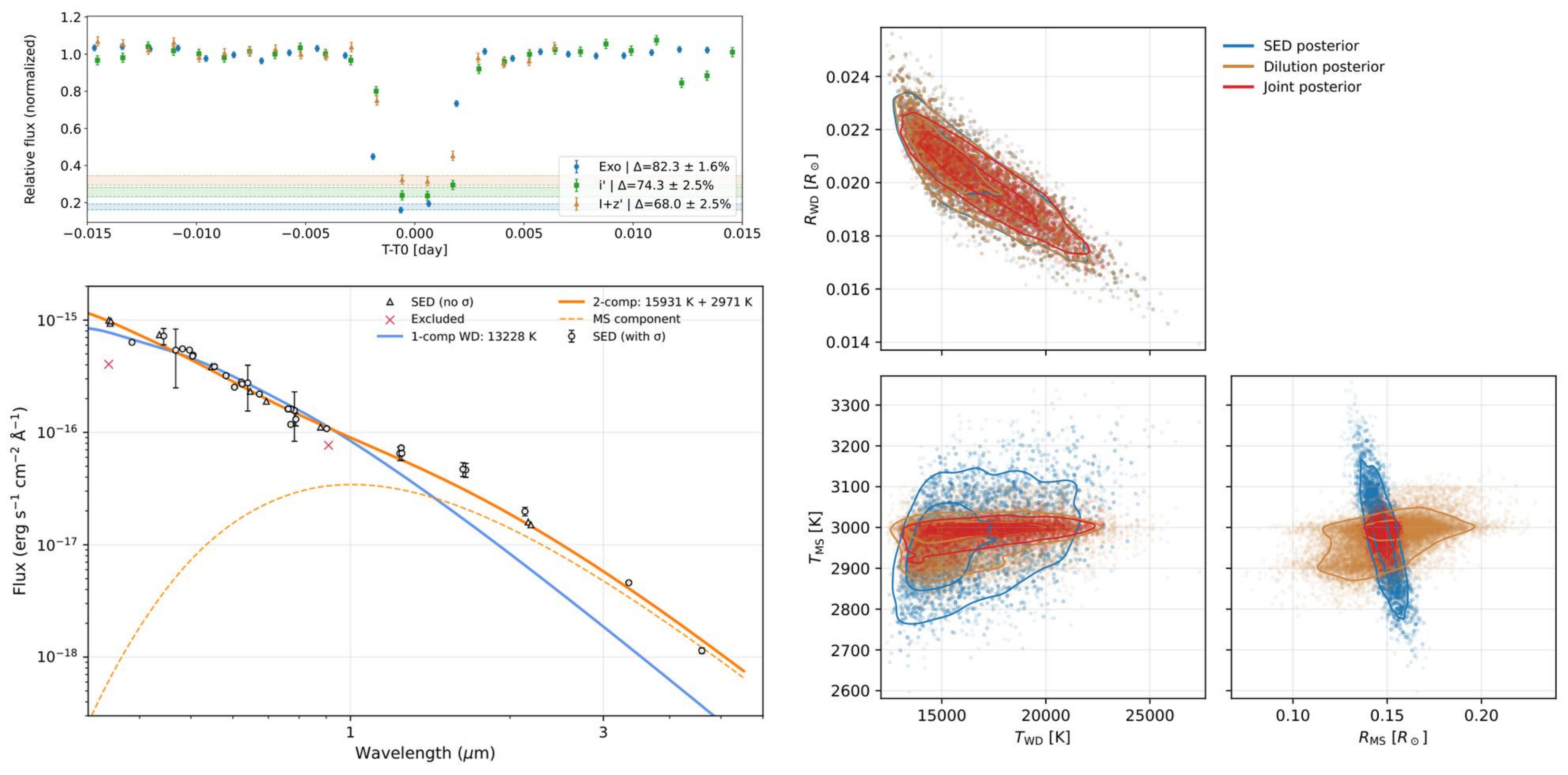}
    \hfill
    \caption{\textbf{Summary figure for TIC450781262's analysis}.
    \textit{Top left}: Follow-up observations obtained with the \textit{SPECULOOS} network of 1-m telescopes showcasing detectable level of transit dilution pointing to eclipses by a M-dwarf companion.
    \textit{Bottom left}: SED showcasing flux excess beyond 1 microns, pointing to a cool-M companion. 
    \textit{Right}: Posterior probability of the key system parameters from transit-only (orange), SED-only (blue), and joint (red) fits showing good agreement between the complementary datasets.
    }
    \label{fig:Fig4507}
\end{figure}

\begin{figure}[!htbp]
    \centering
    \includegraphics[width=1\textwidth]{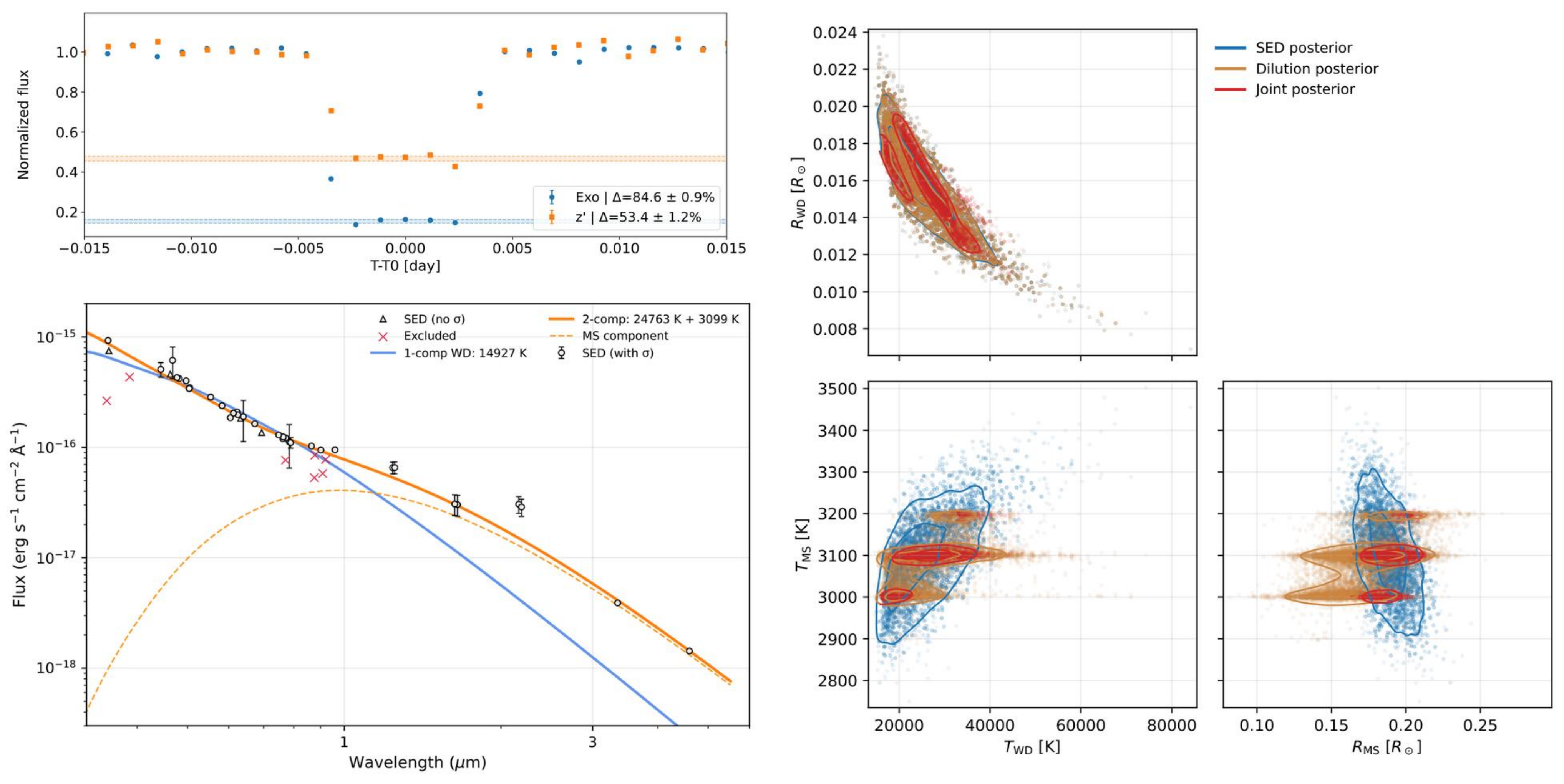}
    \caption{\textbf{Summary figure for TIC53206761's analysis}.
    \textit{Top left}: Follow-up observations obtained with the \textit{SPECULOOS} network of 1-m telescopes showcasing detectable level of transit dilution pointing to eclipses by a M-dwarf companion.
    \textit{Bottom left}: SED showcasing flux excess beyond 1 microns, pointing to a cool-M companion. 
    \textit{Right}: Posterior probability of the key system parameters from transit-only (orange), SED-only (blue), and joint (red) fits showing good agreement between the complementary datasets.}
    \label{fig:Fig5320}
\end{figure}

\begin{figure}[!htbp]
    \centering
    \includegraphics[width=1\textwidth]{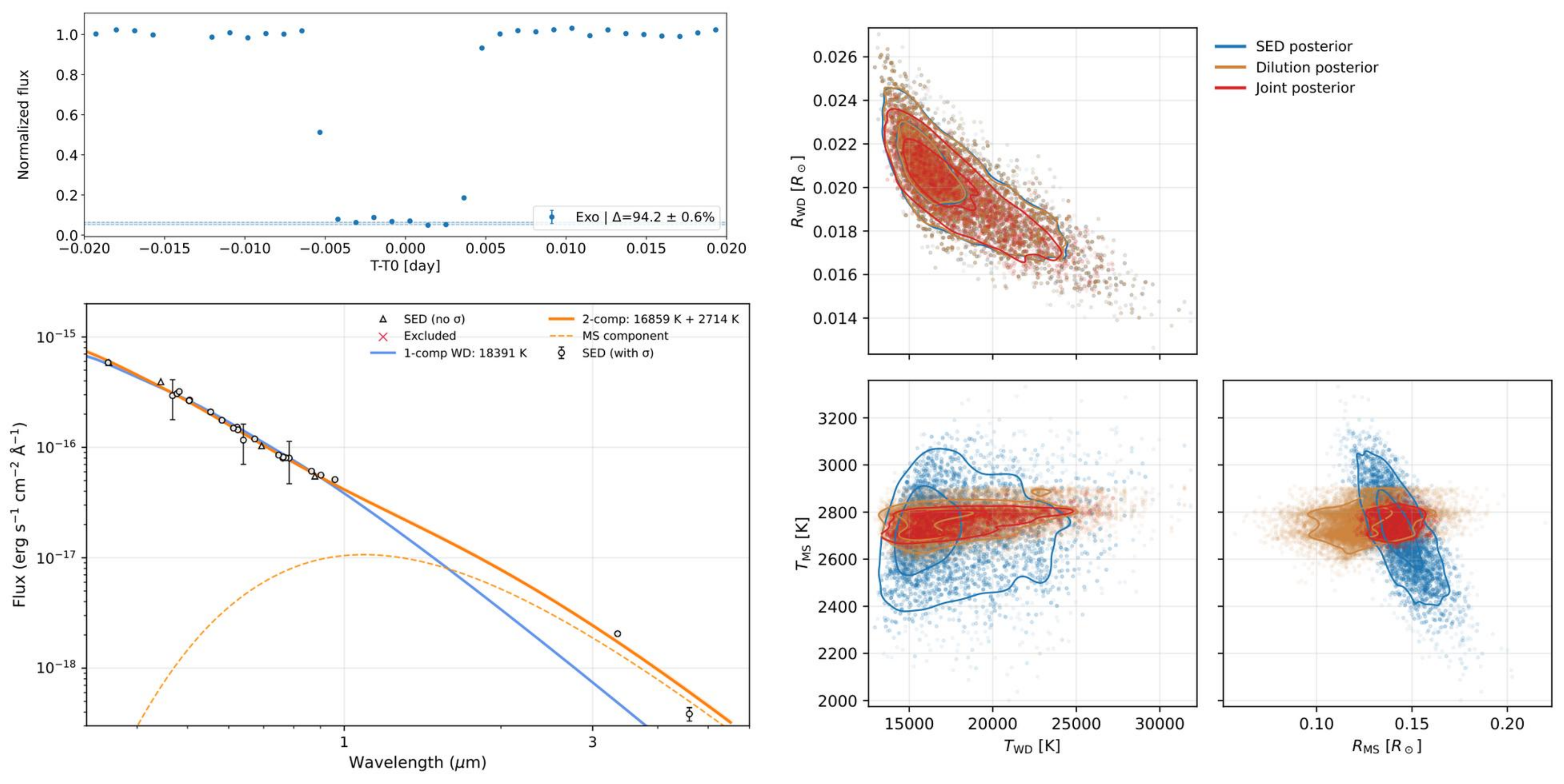}
    \caption{\textbf{Summary figure for TIC2041210548's analysis}.
    \textit{Top left}: Follow-up observations obtained with the \textit{SPECULOOS} network of 1-m telescopes showcasing detectable level of transit dilution pointing to eclipses by a M-dwarf companion.
    \textit{Bottom left}: SED showcasing flux excess beyond 1 microns, pointing to a cool-M companion. 
    \textit{Right}: Posterior probability of the key system parameters from transit-only (orange), SED-only (blue), and joint (red) fits showing good agreement between the complementary datasets.}
    \label{fig:Fig2041}
\end{figure}

\begin{figure}[!htbp]
    \centering
    \includegraphics[width=1\textwidth]{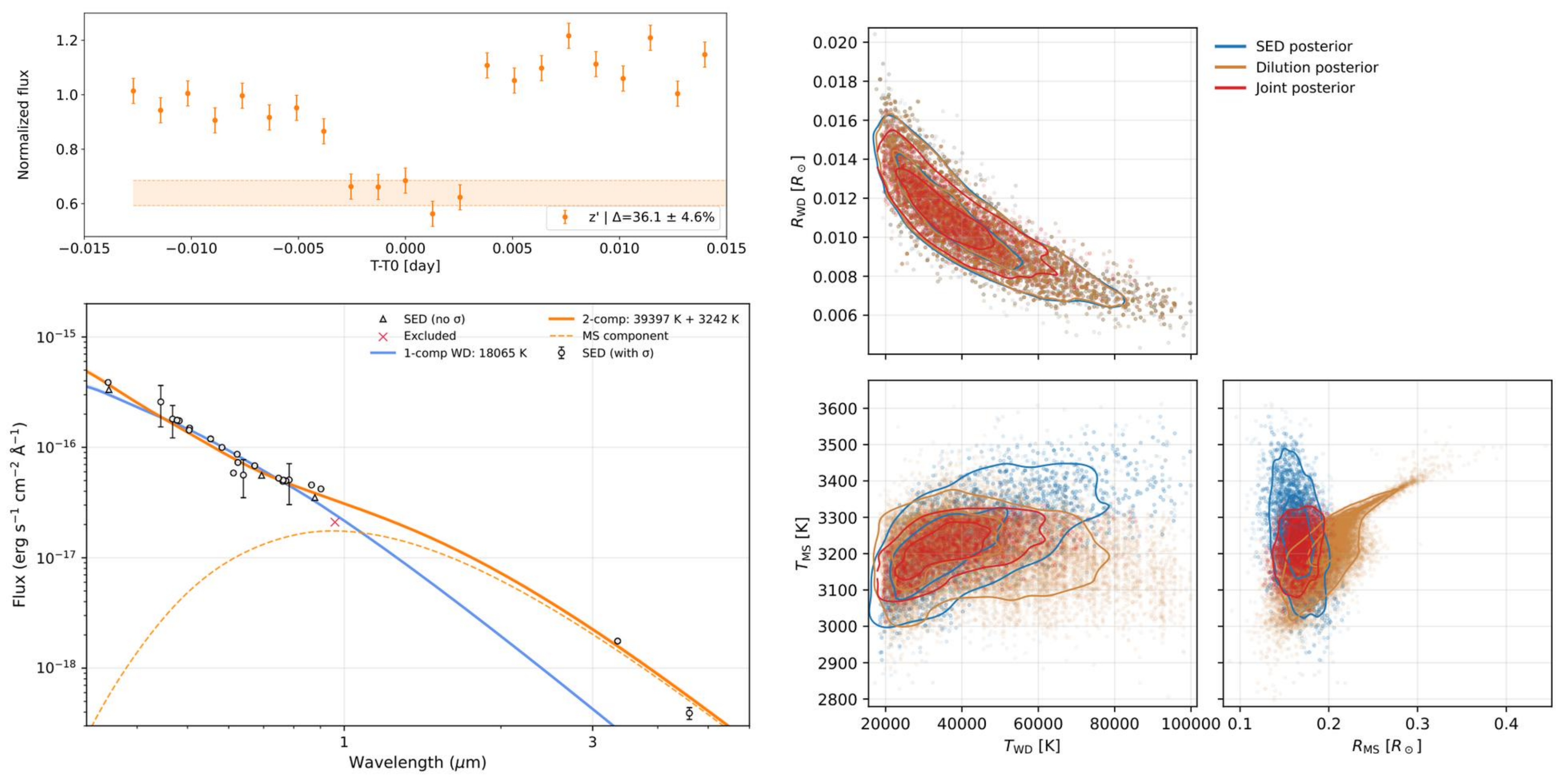}
    \caption{\textbf{Summary figure for TIC1000635787's analysis}.
    \textit{Top left}: Follow-up observations obtained with the \textit{SPECULOOS} network of 1-m telescopes showcasing detectable level of transit dilution pointing to eclipses by a M-dwarf companion.
    \textit{Bottom left}: SED showcasing flux excess beyond 1 microns, pointing to a cool-M companion. 
    \textit{Right}: Posterior probability of the key system parameters from transit-only (orange), SED-only (blue), and joint (red) fits showing good agreement between the complementary datasets.}
    \label{fig:Fig1000}
\end{figure}

\begin{figure}[!htbp]
    \centering
    \includegraphics[width=1\textwidth]{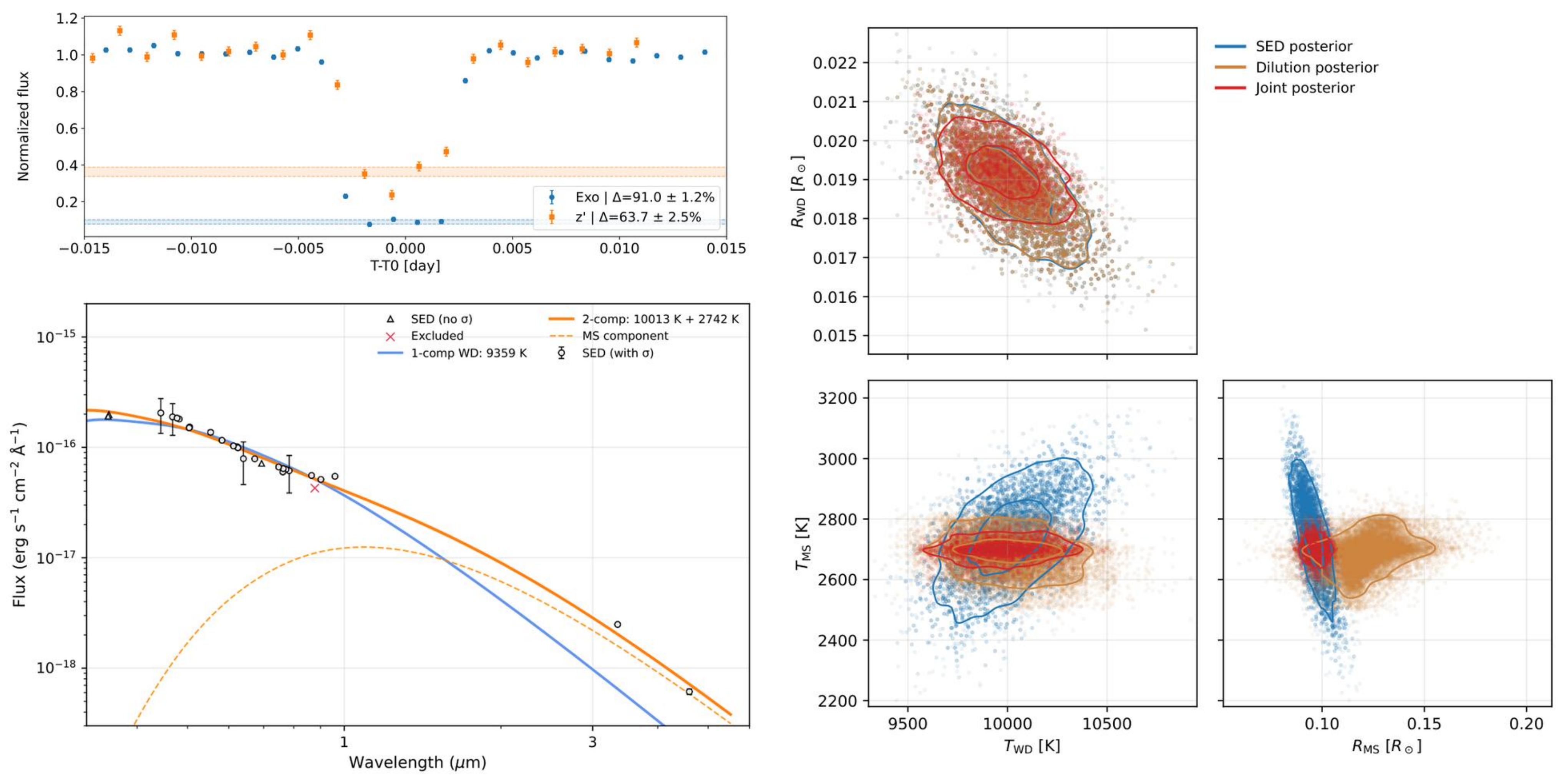}
    \caption{\textbf{Summary figure for TIC1883504789's analysis}.
    \textit{Top left}: Follow-up observations obtained with the \textit{SPECULOOS} network of 1-m telescopes showcasing detectable level of transit dilution pointing to eclipses by a M-dwarf companion.
    \textit{Bottom left}: SED showcasing flux excess beyond 1 microns, pointing to a cool-M companion. 
    \textit{Right}: Posterior probability of the key system parameters from transit-only (orange), SED-only (blue), and joint (red) fits showing good agreement between the complementary datasets.}
    \label{fig:Fig1883}
\end{figure}

\begin{figure}[!htbp]
    \centering
    \includegraphics[width=1\textwidth]{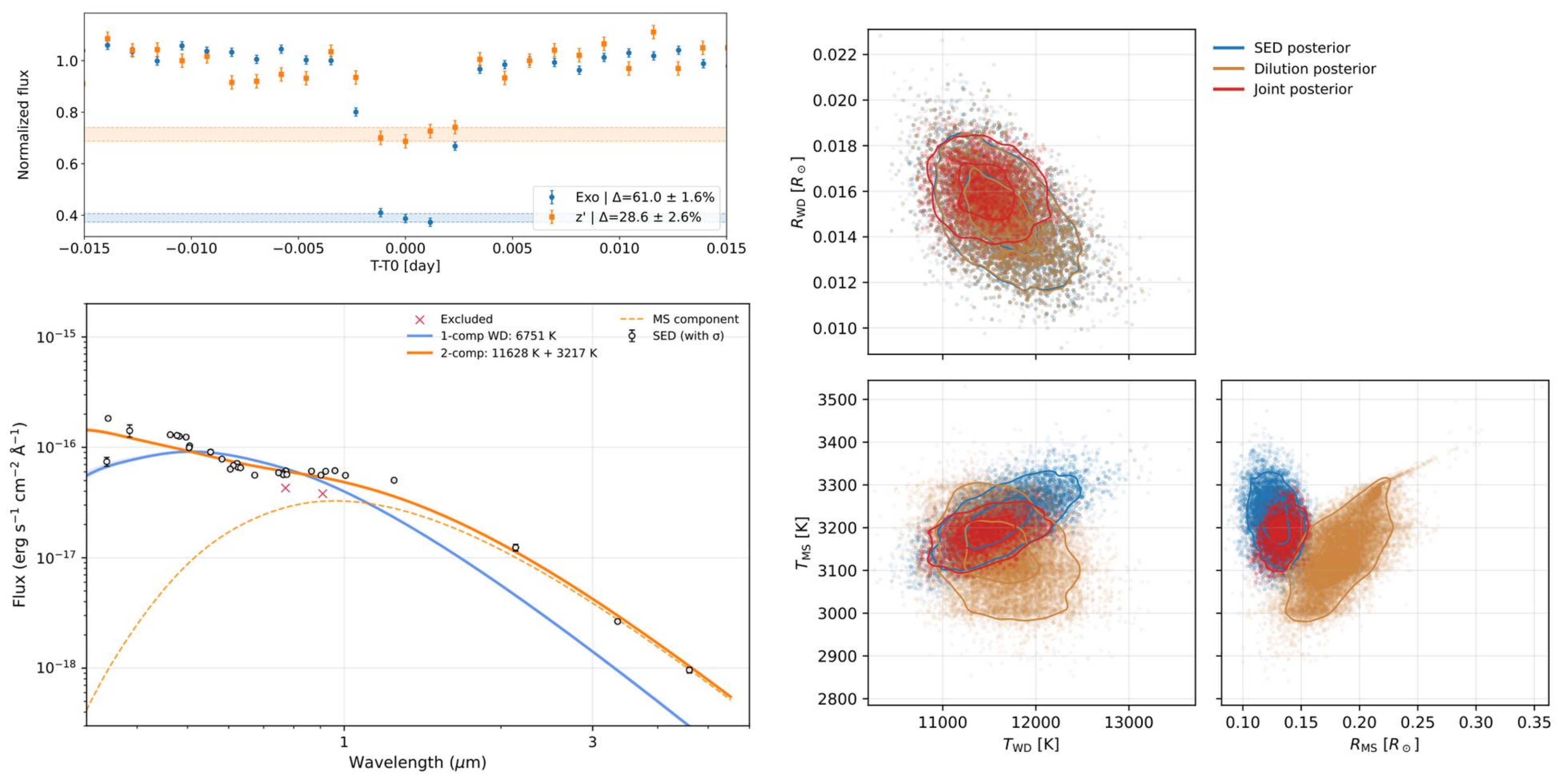}
    \hfill
    \caption{\textbf{Summary figure for TIC610682128's analysis}.
    \textit{Top left}: Follow-up observations obtained with the \textit{SPECULOOS} network of 1-m telescopes showcasing detectable level of transit dilution pointing to eclipses by a M-dwarf companion.
    \textit{Bottom left}: SED showcasing flux excess beyond 1 microns, pointing to a cool-M companion. 
    \textit{Right}: Posterior probability of the key system parameters from transit-only (orange), SED-only (blue), and joint (red) fits showing good agreement between the complementary datasets.}
    \label{fig:Fig6106}
\end{figure}

\begin{figure}[!htbp]
    \centering
    \includegraphics[width=1\textwidth]{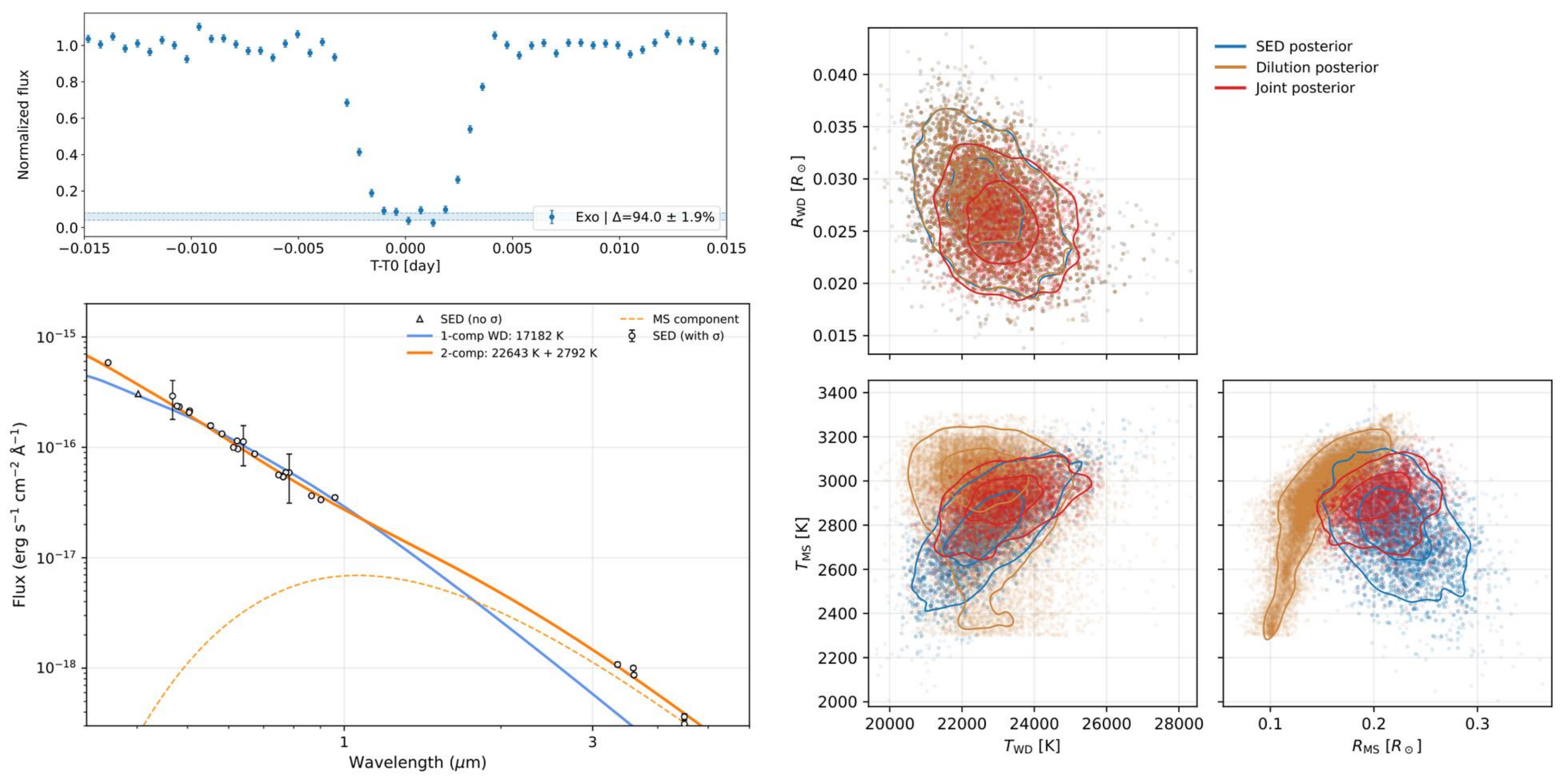}
    \caption{\textbf{Summary figure for TIC600854869's analysis}.
    \textit{Top left}: Follow-up observations obtained with the \textit{SPECULOOS} network of 1-m telescopes showcasing detectable level of transit dilution pointing to eclipses by a M-dwarf companion.
    \textit{Bottom left}: SED showcasing flux excess beyond 1 microns, pointing to a cool-M companion. 
    \textit{Right}: Posterior probability of the key system parameters from transit-only (orange), SED-only (blue), and joint (red) fits showing good agreement between the complementary datasets.}
    \label{fig:Fig6008}
\end{figure}

\begin{figure}[!htbp]
    \centering
    \includegraphics[width=1\textwidth]{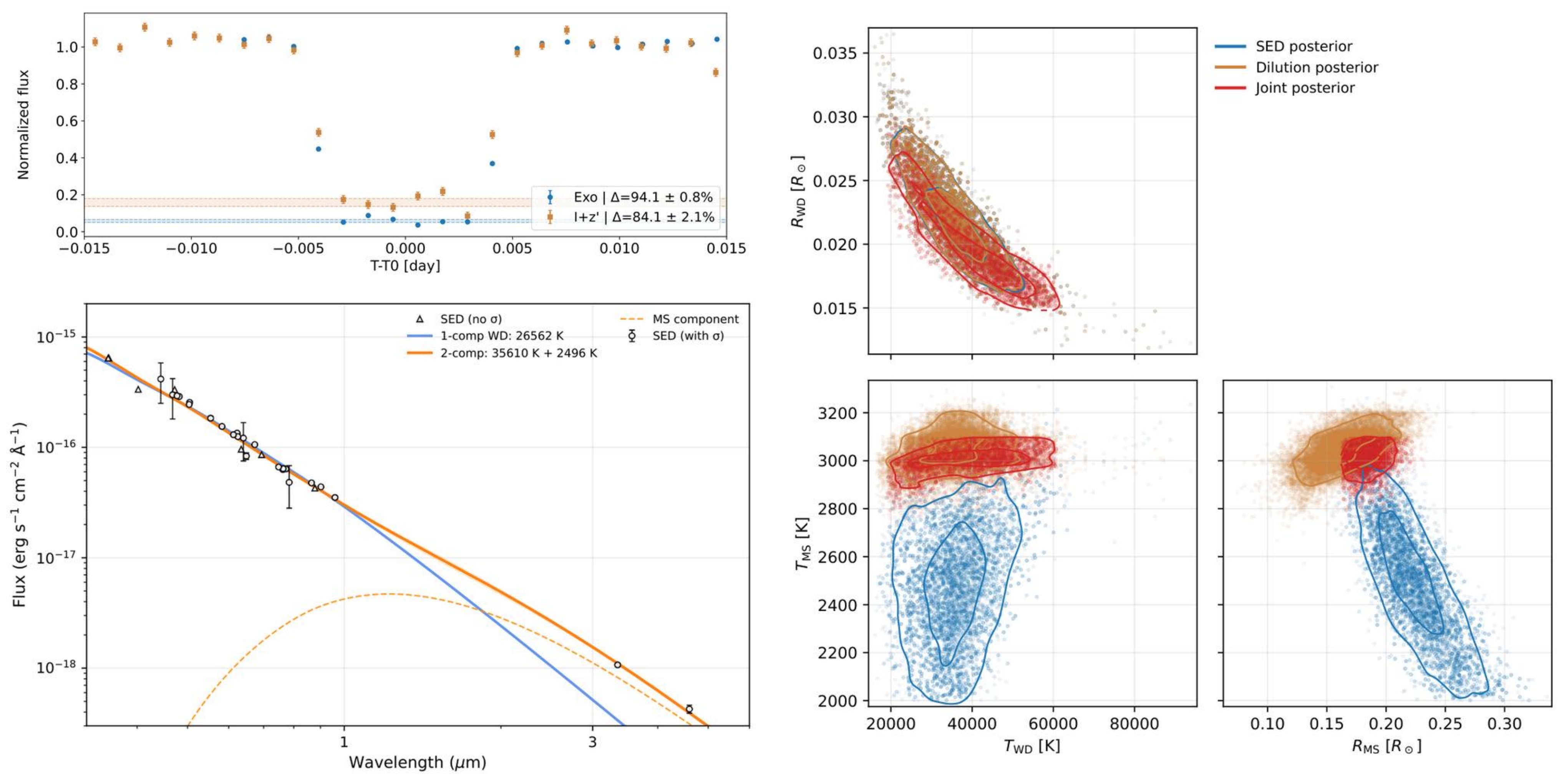}
    \hfill
    \caption{
    \textbf{Summary figure for TIC1201294971's analysis}.
    \textit{Top left}: Follow-up observations obtained with the \textit{SPECULOOS} network of 1-m telescopes showcasing detectable level of transit dilution pointing to eclipses by a M-dwarf companion.
    \textit{Bottom left}: SED showcasing flux excess beyond 1 microns, pointing to a cool-M companion. 
    \textit{Right}: Posterior probability of the key system parameters from transit-only (orange), SED-only (blue), and joint (red) fits showing good agreement between the complementary datasets.}
    \label{fig:Fig1201}
\end{figure}

\begin{figure}[!htbp]
    \centering
    \includegraphics[width=1\textwidth]{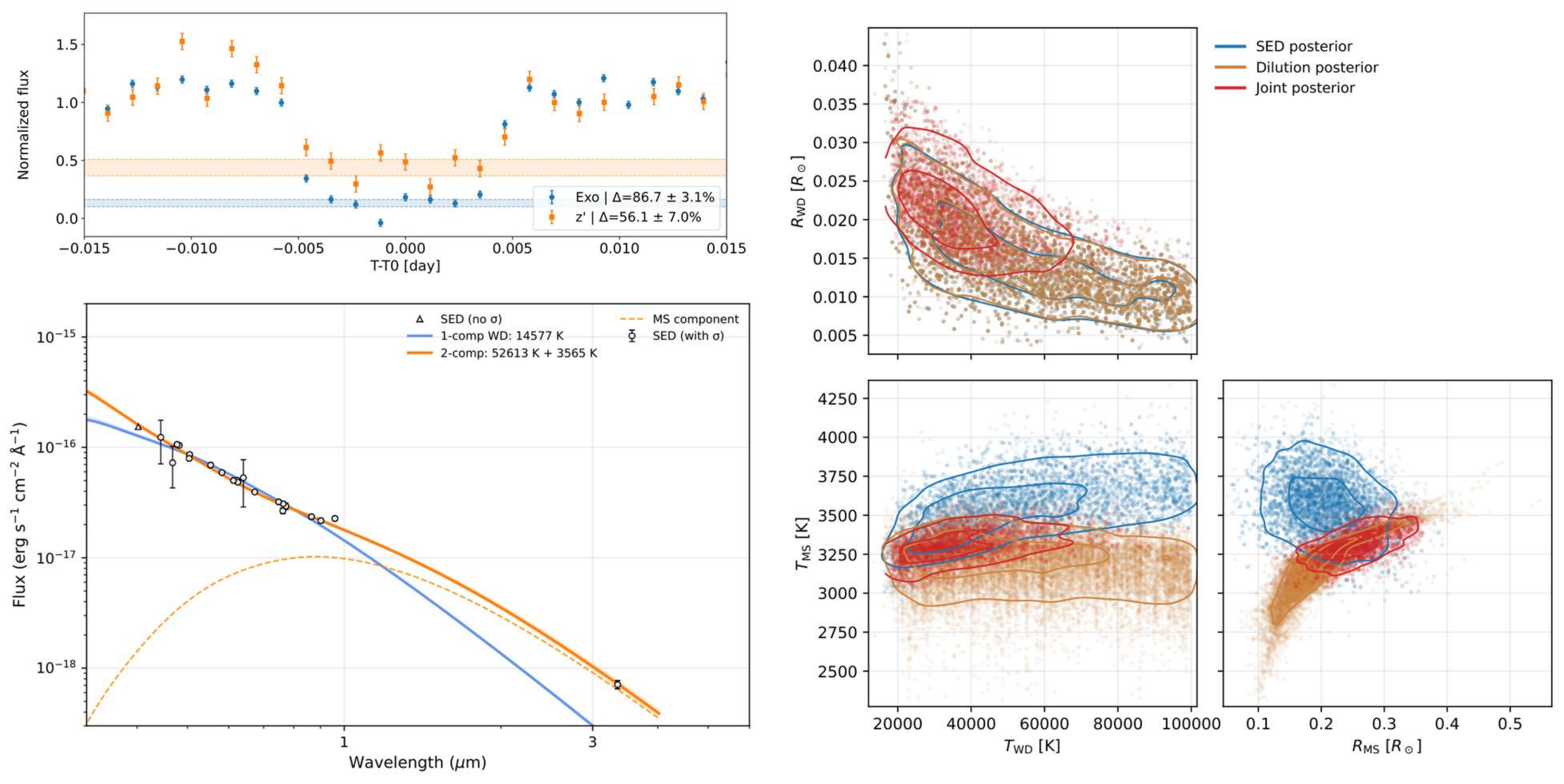}
    \caption{\textbf{Summary figure for TIC1936725512's analysis}.
    \textit{Top left}: Follow-up observations obtained with the \textit{SPECULOOS} network of 1-m telescopes showcasing detectable level of transit dilution pointing to eclipses by a M-dwarf companion.
    \textit{Bottom left}: SED showcasing flux excess beyond 1 microns, pointing to a cool-M companion. 
    \textit{Right}: Posterior probability of the key system parameters from transit-only (orange), SED-only (blue), and joint (red) fits showing good agreement between the complementary datasets.}
    \label{fig:Fig1936}
\end{figure}

\begin{figure}[!htbp]
    \centering
    \includegraphics[width=1\textwidth]{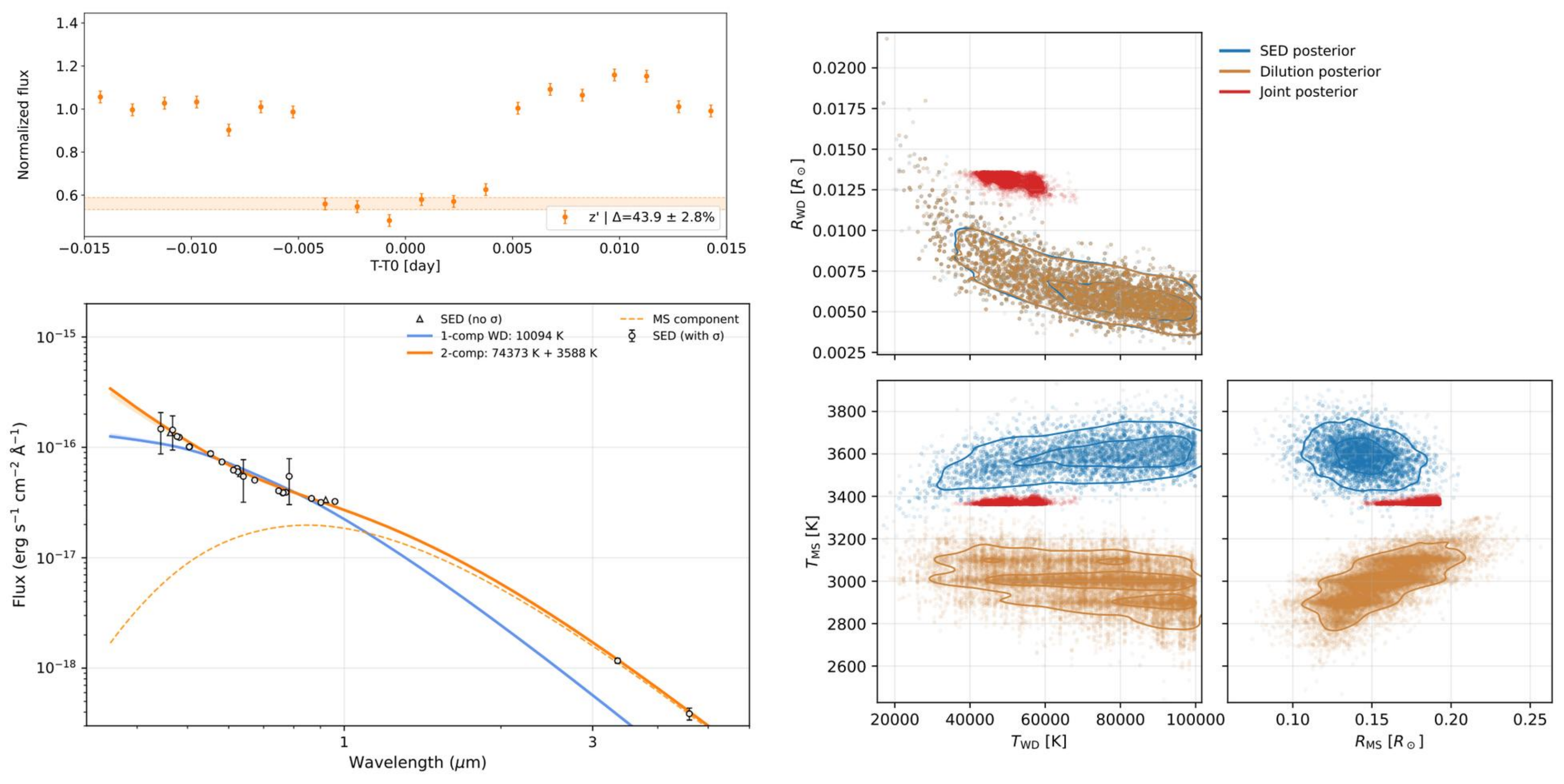}
    \caption{\textbf{Summary figure for TIC1506668467's analysis}.
    \textit{Top left}: Follow-up observations obtained with the \textit{SPECULOOS} network of 1-m telescopes showcasing detectable level of transit dilution pointing to eclipses by a M-dwarf companion.
    \textit{Bottom left}: SED showcasing flux excess beyond 1 microns, pointing to a cool-M companion. 
    \textit{Right}: Posterior probability of the key system parameters from transit-only (orange), SED-only (blue), and joint (red) fits showing good agreement between the complementary datasets.}
    \label{fig:Fig1506}
\end{figure}
\end{document}